\documentclass[journal]{IEEEtran}
\usepackage{amsmath,amsfonts}
\usepackage{algorithmic}
\usepackage{algorithm}
\usepackage{array}
\usepackage[caption=false,font=normalsize,labelfont=sf,textfont=sf]{subfig}
\usepackage{textcomp}
\usepackage{stfloats}
\usepackage{url}
\usepackage{verbatim}
\usepackage{graphicx}
\usepackage{cite}
\usepackage{amssymb}
\begin{document}

\title{Hybrid Beamforming Design for Bistatic Integrated Sensing and Communication Systems}

\author{Tianhao Mao, \IEEEmembership{Student Member,~IEEE}, 
        Jie Yang, \IEEEmembership{Member,~IEEE}, 
        Le Liang, \IEEEmembership{Member,~IEEE}, \\
        and Shi Jin, \IEEEmembership{Fellow,~IEEE}
\thanks{T. Mao, L. Liang, and S. Jin are with the National Mobile Communications Research Laboratory and Frontiers Science Center for Mobile Information Communication and Security, Southeast University, Nanjing 210096, China (e-mail: tianhao@seu.edu.cn; yangjie@seu.edu.cn; lliang@seu.edu.cn; jinshi@seu.edu.cn). L. Liang is also with the Purple Mountain Laboratories, Nanjing 211111, China.}
\thanks{J. Yang is with the Frontiers Science Center for Mobile Information Communication and Security, Southeast University, Nanjing 210096, China, and also with the Key Laboratory of Measurement and Control of Complex Systems of Engineering, Ministry of
Education, Southeast University, Nanjing 210096, China (e-mail: yangjie@seu.edu.cn).}
}



\maketitle

\begin{abstract}
Integrated sensing and communication (ISAC) in millimeter wave is a key enabler for next-generation networks, which leverages large bandwidth and extensive antenna arrays, benefiting both communication and sensing functionalities. The associated high costs can be mitigated by adopting a hybrid beamforming structure. However, the well-studied monostatic ISAC systems face challenges related to full-duplex operation. To address this issue, this paper focuses on a three-dimensional bistatic configuration that requires only half-duplex base stations. To intuitively evaluate the error bound of bistatic sensing using orthogonal frequency division multiplexing waveforms, we propose a positioning scheme that combines angle-of-arrival and time-of-arrival estimation, deriving the closed-form expression of the position error bound (PEB). Using this PEB, we develop two hybrid beamforming algorithms for joint waveform design, aimed at maximizing achievable spectral efficiency (SE) while ensuring a predefined PEB threshold. The first algorithm leverages a Riemannian trust-region approach, achieving superior performance in terms of global optima and convergence speed compared to conventional gradient-based methods, but with higher complexity. In contrast, the second algorithm, which employs orthogonal matching pursuit, offers a more computationally efficient solution, delivering reasonable SE while maintaining the PEB constraint. Numerical results are provided to validate the effectiveness of the proposed designs.
\end{abstract}


\section{Introduction}

\IEEEPARstart{I}{ntegrated} sensing and communication (ISAC) is a key technology for next-generation wireless networks \cite{liufov}. With the convergence of radar and communication systems \cite{mimoradar}, ISAC enhances resource efficiency and offers mutual benefits, enabling innovative applications across various industries \cite{andrewov}. The millimeter wave (mmWave) spectrum, with its wide bandwidth and substantial beamforming gains, makes mmWave-based ISAC a promising direction for future networks.

The short wavelength of mmWave signals allows base stations (BS) to deploy large-scale antenna arrays; however, fully digital beamforming is impractical due to its high cost. Hybrid beamforming for mmWave multiple-input multiple-output (MIMO) systems has been proposed, which reduces the required number of radio frequency (RF) chains by interfacing a low-dimensional digital beamformer with a high-dimensional analog beamformer \cite{overview}, \cite{lintian}. For point-to-point MIMO, a common approach frames hybrid beamforming as a matrix factorization problem, minimizing the Euclidean distance between the hybrid and fully digital beamformers \cite{spatially}-\cite{alternating}. First introduced in \cite{spatially}, this design leverages mmWave spatial characteristics and uses orthogonal matching pursuit (OMP) to approximate spectral efficiency (SE) maximization. The authors in \cite{alternating} further propose a manifold optimization (MO)-based algorithm, addressing the unit-modulus constraint on a Riemannian manifold for both narrowband and wideband cases. Beyond matrix factorization, the work in \cite{yuwei} and \cite{yuwei2} tackle SE maximization directly, offering closed-form solutions and iterative algorithms.

The aforementioned works focus only on SE, addressing communication-only designs. However, for ISAC joint waveform design, both communication and sensing performance must be considered, significantly increasing the complexity of the problem. To realize simultaneous communication and monostatic sensing via hybrid beamforming in a dual-functional radar-communication (DFRC) BS, the authors in \cite{hbficassp} innovatively apply hybrid beamforming structures to ISAC beamforming design, successfully approximating the optimal communication and sensing beamformers simultaneously. The work in \cite{partially} examines the beamforming design for a DFRC BS with a partially-connected structure, where the problem is formulated as minimizing the Cramér-Rao bound (CRB) of a single target's angle of departure (AOD) while ensuring a pre-defined level of communication signal-to-interference-plus-noise ratio (SINR) for each single-antenna user. This model is extended to multi-target scenarios in \cite{gong}, where the weighted minimum mean square error approach is used to address the weighted sum-rate problem, with CRB constrained under a specified level. Both \cite{partially} and \cite{gong} solve the problem via MO with gradient-based algorithms such as steepest descent and conjugate gradient descent. In wideband cases with orthogonal frequency division multiplexing (OFDM) signals, the work in \cite{hbfofdm} formulates the objective as a weighted sum of SE for multiple users and spatial spectrum matching error for radar, solving the problem via the alternating direction method of multipliers (ADMM) algorithm.

However, in monostatic sensing configurations, managing self-interference becomes critical when a monostatic BS operates in full-duplex mode, which introduces additional resource consumption \cite{fd2}. To handle full-duplex issues and leverage cooperative sensing in multi-static scenarios, beamforming design for multi-static configurations has been explored. \cite{bistatic1} consider a bistatic scenario and propose to design the transmit signal matrix to minimize the CRB of the angle of arrival (AOA) and AOD while guarantee the SINR of the users. The problem is formulated as a quartic optimization problem, and solved via ADMM algorithm. To achieve seamless sensing coverage, the work in \cite{seamless} investigate the communication SINR and sensing signal-to-noise ratio (SNR), starting by beamforming design in bistatic settings and extend it into multi-static case. However, the studies above focus only on the single-carrier narrowband case and do not consider the hybrid beamforming structure.

In this paper, we derive an intuitive metric, the position error bound (PEB), for bistatic sensing, based on which the joint hybrid beamforming in the MIMO-OFDM system is performed. To avoid self-interference in monostatic configuration, bistatic sensing, also known as passive sensing, is adopted. To the best of our knowledge, this study is the first to discuss joint hybrid beamforming design that integrates passive sensing and communication in mmWave MIMO-OFDM systems. Unlike recent studies on ISAC beamforming design which assume other parameters are known when deriving error bounds for the parameters of interest \cite{partially}, \cite{gong}, we compute PEB for an unknown sensing channel. Our main contributions are presented in detail as follows:
\begin{itemize}
    \item We derive the FIM for channel parameter estimation using OFDM waveforms, and further develop the work in \cite{henk} to prove that under extensive receive arrays, the estimation of azimuth and elevation AOA is independent of each other. We then formulate the FIM of AOA and time of arrival (TOA) estimation as a $3\times3$ diagonal matrix, indicating that the estimation of azimuth (elevation) AOA and TOA is independent of other parameters, with complex channel gain uncertainty impacting only AOD estimation. Given that in a bistatic ISAC system, the positions of the transmit and receive BSs are perfectly known, we propose target positioning via the hybrid AOA/TOA positioning scheme and derive the corresponding closed-form PEB in three-dimentional (3D) space. For target positioning, unlike \cite{bistatic1} and \cite{bistatic2} which express the error bound as a fractional quartic function with respect to (w.r.t) the beamformer, we find it to be quadratic, simplifying the problem significantly.

    \item Based on the alternating minimization approach, we optimize the analog and digital beamformers alternatively. For the analog beamformer, the widely used gradient-based algorithms in \cite{alternating}, \cite{partially}-\!\cite{gong} rely only on first-order information, often suffering from local optima and slow convergence. To address this, we incorporate second-order information, represented by the Riemannian Hessian matrix, and apply the Riemannian trust-region (RTR) algorithm. For the digital beamformer, we employ successive convex approximation (SCA) to address the non-convex nature of the problem. Numerical results illustrate that the integrated RTR-SCA algorithm effectively finds the global optimal and accelerates the convergence.
    
    \item Leveraging the orthogonality of different steering vectors in large antenna arrays, we propose an OMP-based algorithm for positioning and communication (PC-OMP) as a low-complexity alternative to the RTR-SCA algorithm. Specifically, we decompose the hybrid beamformer into two components: one dedicated to positioning and the other to communication. These two components are shown to be independent, allowing for separate design optimizations for each function without interference.
\end{itemize}

The rest of the paper is organized as follows. We first introduce the system model, including the transmission model and channel models in Section \ref{sec:systemmodel}. Next, in Section \ref{sec:metric}, the performance metrics for communication and postioning are respectively derived, based on which the problem is formulated. A hybrid beamforming algorithm for fully-connected structure is then proposed in Section \ref{sec:mo} to solve the optimization problem. In Section \ref{sec:omp}, another low-complexity algorithm named PC-OMP is demonstrated. Simulation results are presented in Section \ref{sec:simulation}. Finally, we conclude the paper in Section \ref{sec:conclusion}.

The notations used are defined as follows. $\mathbf{a}$ and $\mathbf{A}$ denote a column vector and a matrix, respectively. $(\mathbf{A})_{i,j}$ refers to the entry at the $i$-th row and $j$-th column of $\mathbf{A}$. The conjugate, transpose, and conjugate transpose of $\mathbf{A}$ are denoted by $\mathbf{A}^*$, $\mathbf{A}^{\mathrm{T}}$, and $\mathbf{A}^{\mathrm{H}}$, respectively. The determinant and trace of $\mathbf{A}$ are given by $\mathrm{det}(\mathbf{A})$ and $\mathrm{tr}(\mathbf{A})$, and $\|\mathbf{A}\|_F$ denotes the Frobenius norm. Expectation and absolute value of a complex variable are denoted by $\mathbb{E}[\cdot]$ and $|\cdot|$, respectively. A complex variable's real and imaginary parts are represented by $\Re\{\cdot\}$ and $\Im\{\cdot\}$. $\mathrm{tr}(\mathbf{A})$ and $\mathrm{vec}(\mathbf{A})$ denote the trace and vectorization of $\mathbf{A}$. The Hadamard and Kronecker products of two matrices are represented by $\odot$ and $\otimes$, respectively. $\mathrm{max}\{a,b\}$ denotes the maximum value between $a$ and $b$ and $\mathrm{min}\{a,b\}$ stands for the minimum value. The inner product is denoted by $\left\langle\cdot,\cdot\right\rangle$. 

\section{System Model} \label{sec:systemmodel}
\begin{figure}[t] \vspace{-0.4cm}
	\centerline{\includegraphics[width=3in]{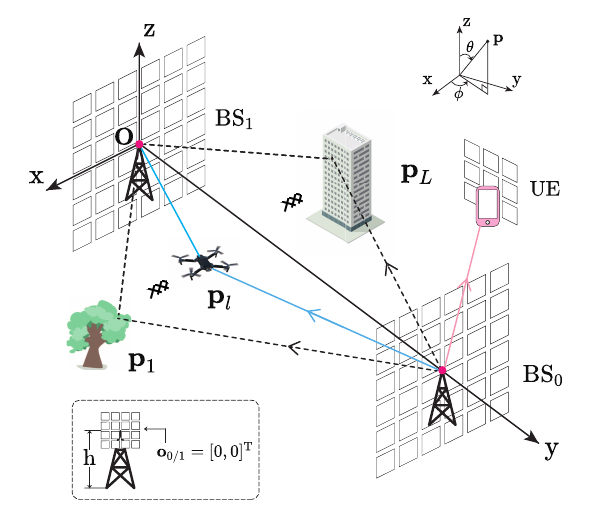}}
	\caption{The DFRC system in bistatic configuration.}
	\label{fig:system}
\end{figure}

Consider the mmWave bistatic system with multiple scatterers in the environment, as shown in Fig. \ref{fig:system}. A DFRC transmit BS, $\text{BS}_0$, with $N_{\mathrm{t}}$ antennas performs the single-user MIMO communication and point target positioning simultaneously with the assistance of the DFRC receive BS, $\text{BS}_1$, with $N_{\mathrm{r}}^{\mathrm{s}}$ antennas. Both BSs of the height $h$ are equipped with uniform square planar arrays (USPA), with their geometric centers positioned at $(0, D, 0)$ and $(0, 0, 0)$, respectively. Here, $D$ represents the distance between the two BSs. Both $D$ and $h$ are measured in meters. For simplicity, the orientation angles of the USPA at $\text{BS}_0$, denoted by $\mathbf{o}_0$, and at $\text{BS}_1$, denoted by $\mathbf{o}_1$, are assumed to be $[0, 0]^{\mathrm{T}}$. Notably, the point target positioning is performed with unknown clusters in the sensing channel.

\subsection{Transmission Model} 

We consider an OFDM system with $K$ subcarriers, each spaced by $\Delta f$ in the frequency domain, where the total bandwidth is given by $B = K\Delta f$. During the transmission, the $\text{BS}_0$ transmits the random signal 
$\mathbf{S}_{k}\in\mathbb{C}^{N_{\mathrm{s}}\times M}$ at the $k$-th subcarrier, 
which constitutes $N_{\mathrm{s}}$ data streams and $M$ symbols per stream.

The $\text{BS}_0$ is equipped with $N_{\mathrm{RF}}$ transmit chains, such that $N_{\mathrm{s}}\leq N_{\mathrm{RF}}\leq N_{\mathrm{t}}$. Consequently, beamforming is realized through an $N_{\mathrm{t}}\times N_{\mathrm{RF}}$ RF beamformer, $\mathbf{F}_{\mathrm{RF}}$, and an $N_{\mathrm{RF}}\times N_{\mathrm{s}}$ digital beamformer, 
$\mathbf{F}_{\mathrm{BB},k}$. The signal at the $k$-th subcarrier after beamforming can thus be written as
\begin{equation} \vspace{-0.1cm}
    \mathbf{X}_k = \sqrt{E_0}\mathbf{F}_{\mathrm{RF}}\mathbf{F}_{\mathrm{BB},k}\mathbf{S}_k,
\end{equation}
where $E_0$ is the energy per symbol duration and $\mathbf{S}_{k}=\big[\mathbf{s}_k[1],...,\mathbf{s}_k[M]\big]$. Without loss of generality, we assume $\mathbb{E}[\mathbf{s}_k\mathbf{s}_k^\mathrm{H}]=\frac{1}{N_\mathrm{s}}\mathbf{I}_{N_\mathrm{s}}$ and $\|\mathbf{F}_{{\mathrm{RF}}}\mathbf{F}_{{\mathrm{BB},k}}\|_{F}^{2}=N_{{\mathrm{s}}}$. Furthermore, given that the analog beamformer is implemented with phase shifters, the entries of $\mathbf{F}_{{\mathrm{RF}}}$ should satisfy the unit-modulus constraint, that is, $|(\mathbf{F}_{{\mathrm{RF}}})_{i,j}|=1, \forall i,j$.

We note that the random signal after beamforming serves both communication and target positioning. For the communication receiver UE, the signal is random and conveys information. For the sensing receiver $\text{BS}_1$, the signal is random but known. After scattering off the targets of interest (TOIs) and other objects, $\text{BS}_1$ can extract sensing parameters from the scattered signal. To extend to a multi-target scenario, the targets can be positioned block-by-block.

\subsection{Communication Channel}
Due to the high free-space path loss during mmWave propagation and highly correlated antennas in mmWave transceivers, a clustered channel model, i.e., the Saleh-Valenzuela model \cite{svchannel} is adopted. Assuming the channel has $N_{\mathrm{cl}}$ clusters, each of which contributes $N_{\mathrm{ray}}$ paths to the channel matrix,
the model depicts the mmWave channel matrix of the $k$-th subcarrier in frequency domain as
\begin{equation} \vspace{-0.1cm}
    \mathbf{H}_{k}^{\mathrm{c}}=\gamma_0\sum_{i=1}^{N_\mathrm{cl}}\sum_{j=1}^{N_\mathrm{ray}}\alpha_{ij}\mathbf{b}_{ij}(\theta_{ij}^\mathrm{r},\phi_{ij}^\mathrm{r})\mathbf{a}^\mathrm{H}_{ij}(\theta_{ij}^\mathrm{t},\phi_{ij}^\mathrm{t})e^{-j2\pi f_k\tau_i}.
\end{equation}
Here, $\tau_i$ denotes the delay of the $i$-th cluster. The frequency $f_k$ satisfies $f_k=(k-\frac{K+1}{2})\Delta f$, with $k=1,...,K$. $\alpha_{ij}$ is the complex gain of the $l$-th path of the $i$-th cluster, following the complex Gaussian distribution $\mathcal{CN}(0,\sigma_{\alpha,i}^2)$, where $\sigma_{\alpha,i}^2$ represents the average power of the $i$-th cluster. The sum of the average cluster power satisfies $\sum_{i=1}^{N_\mathrm{cl}}\sigma_{\alpha,i}^2=\gamma_0$, which normalize the channel such that $\mathbb{E}\left[\left\|\mathbf{H}_{k}^{\mathrm{c}}\right\|_F^2\right]=N_{\mathrm{t}}N_{\mathrm{r}}^{\mathrm{c}}$, with $N_{\mathrm{r}}^{\mathrm{c}}$ denoting the number of receive antennas in UE. In addition, $\phi_{ij}^r(\theta_{ij}^r)$ and $\phi_{ij}^\mathrm{t}(\theta_{ij}^\mathrm{t})$ represent the azimuth (elevation) angle of arrival and departure respectively. Given the USPA settings, the unit-norm steering vectors are written as
\begin{subequations} \label{eq:steeringvector}
   \begin{align}&\mathbf{a}_{ij}(\theta_{ij},\phi_{ij})\triangleq\frac{1}{\sqrt{N_{\mathrm{t}}}}e^{-j\boldsymbol{\Lambda}_{\mathrm{t}}^{\mathrm{T}}\mathbf{k}(\theta_{ij},\phi_{ij})},\\&\mathbf{b}_{ij}(\theta_{ij},\phi_{ij})\triangleq\frac{1}{\sqrt{N_{\mathrm{r}}^{\mathrm{c}}}}e^{-j\boldsymbol{\Lambda}_{\mathrm{r}}^{\mathrm{c}\,\mathrm{T}}\mathbf{k}(\theta_{ij},\phi_{ij})},\end{align} 
\end{subequations}
where $\boldsymbol{\Lambda}_{\mathrm{t}}=[\mathbf{v}_{\mathrm{t},1},\mathbf{v}_{\mathrm{t},2},...,\mathbf{v}_{\mathrm{t},N_{\mathrm{t}}}]\in\mathbb{C}^{3\times N_{\mathrm{t}}}$, $\mathbf{v}_{\mathrm{t},n}=[x_{\mathrm{t},n},y_{\mathrm{t},n},z_{\mathrm{t},n}]^{\mathrm{T}}$ is the Cartesian coordinates of the $n$-th transmitter antenna element, which is assumed half-wavelength spaced. $\boldsymbol{\Lambda}_{\mathrm{r}}^{\mathrm{c}}$ can be similarly defined. The wavenumber vector, $\mathbf{k}(\theta,\phi)$, is given by $\mathbf{k}(\theta,\phi)=\frac{2\pi}\lambda[\sin\theta\cos\phi,\sin\theta\sin\phi,\cos\theta]^\mathrm{T}$, where $\lambda$ is the wavelength.

The received communication signal at the $k$-th subcarrier $\mathbf{Y}_{k}^{\mathrm{c}}$ can be expressed as
\begin{equation} \vspace{-0.1cm}
\mathbf{Y}_{k}^{\mathrm{c}}=\sqrt{E_0}\mathbf{H}_{k}^{\mathrm{c}}\mathbf{F}_{{\mathrm{RF}}}\mathbf{F}_{{\mathrm{BB},k}}\mathbf{S}_{k}+\mathbf{N}_{k}^{\mathrm{c}},
\end{equation}
where $\mathbf{N}_{k}^{\mathrm{c}}\in\mathbb{C}^{N_r\times M}$ is the communication noise matrix, whose entries are identical and independently distributed (i.i.d.) circularly symmetric complex Gaussian random variables with zero mean and a variance of $\sigma_{\mathrm{c}}^2$.

\subsection{Sensing Channel}
We assume there are $L$ scatterers between the $\text{BS}_0$ and the $\text{BS}_1$ located at $\mathbf{p}_l=[x_l,y_l,z_l]^{\mathrm{T}},1\leq l\leq L$ that contributes to $L$ paths of the sensing channel, including the TOIs. There is also a line-of-sight (LOS) path, with $l=0$. Denoting the distance between the $l$-th scatterer and the $\text{BS}_0$ as $d_{0,l}$, and the distance between the $l$-th scatterer and the $\text{BS}_1$ as $d_{1,l}$, the sensing channel for the $k$-th subcarrier in the frequency domain can be written as \cite{yangjie}
\begin{equation} \vspace{-0.1cm}
    \mathbf{H}_{k}^{\mathrm{s}}=\sum_{l=0}^L\mathbf{H}_le^{-j2\pi f_k\tau_l}.
\end{equation}
Assuming that two BSs are synchronized via an optical fiber connection, the TOA of the $l$-th path is given by $\tau_{l}=(d_{0,l}+d_{1,l})/c$, for $l\geq 1$. For the $l$-th path, we have
\begin{equation} \vspace{-0.1cm}
\mathbf{H}_l=\sqrt{N_{\mathrm{t}}N_{\mathrm{r}}^{\mathrm{s}}}\beta_{l}\mathbf{b}_{l}(\theta^{\mathrm{r}}_l,\phi^{\mathrm{r}}_l)\mathbf{a}_{l}^{\mathrm{H}}(\theta^{\mathrm{t}}_l,\phi^{\mathrm{t}}_l),
\end{equation}
where $\mathbf{b}_l$ and $\mathbf{a}_l$ are the receive and transmit steering vector defined similarly by \eqref{eq:steeringvector}. $N_{\mathrm{r}}^{\mathrm{s}}$ represents the number of receive antennas in $\text{BS}_1$ and the complex channel gain is modeled as
\begin{equation} \vspace{-0.1cm}
    \begin{aligned}|\beta_l|^2=\frac{\lambda^2}{(4\pi)^2}\begin{cases}1/D^2&\text{LOS},\\\sigma_{\text{RCS}}^2/(4\pi(d_{0,l}d_{1,l})^2)&\text{scatterer},\end{cases}\end{aligned}
\end{equation}
where $\sigma_{\text{RCS}}^2$ is the radar cross section measured by m$^2$. Thus, the signal received by the $\text{BS}_1$ at the $k$-th subcarrier is expressed as
 \begin{equation} \vspace{-0.1cm} \label{eq:received}
     \mathbf{Y}^{\mathrm{s}}_k=\sqrt{E_0}\mathbf{H}_{k}^{\mathrm{s}}\mathbf{F}_{{\mathrm{RF}}}\mathbf{F}_{{\mathrm{BB},k}}\mathbf{S}_{k}+\mathbf{N}_{k}^{\mathrm{s}},
 \end{equation}
 where $\mathbf{N}_{k}^{\mathrm{s}}$ represents the sensing noise matrix, whose entries are i.i.d. circularly symmetric complex Gaussian random variables with zero mean and a variance of $\sigma_{\mathrm{s}}^2$.
 
\section{Performance Metric} \label{sec:metric}
This section defines the performance metrics for both communication and sensing. Specifically, for communication, the SE achieved using Gaussian signaling is adopted, while for sensing, a hybrid positioning scheme that combines AOA and TOA estimation is proposed, with the closed-form squared-PEB (SPEB) derived. Based on these two performance metrics, the problem is then properly formulated.
\subsection{Achievable SE}
For single-user MIMO communication, the achievable SE is chosen as the performance metric. Notably, instead of designing both the beamformer and the combiner, we decouple the joint transmitter-receiver optimization problem and focus on the transmitter side. The achievable SE with the transmitted signal following a Gaussian distribution can be written as
\begin{equation} \vspace{-0.1cm}
    \mathcal{R}=\frac{1}{K}\sum\nolimits_{k=1}^K\!\mathcal{R}_k,
\end{equation}
with
\begin{equation} \vspace{-0.1cm}
    \mathcal{R}_k\!\!=\!\log_{2}\!\left(\left|\mathbf{I}\!+\!\frac{\text{SNR}_{\mathrm{c}}}{N_{\mathrm{s}}}\mathbf{H}_{k}^{\mathrm{c}}\mathbf{F}_{\mathrm{RF}}\mathbf{F}_{\mathrm{BB},k}\mathbf{F}_{\mathrm{BB},k}^{\mathrm{H}}\mathbf{F}_{\mathrm{RF}}^{\mathrm{H}}\mathbf{H}_{k}^{\mathrm{c}\,\mathrm{H}}\right|\right)\!\!,
\end{equation}
where $\text{SNR}_{\mathrm{c}}$ represents the SNR of communication. In the narrowband case, through several reasonable approximations \cite{spatially}, maximizing the achievable SE is proved equivalent to minimizing the Euclidean distance, $\|\mathbf{F}_{\mathrm{opt}}-\mathbf{F}_\mathrm{RF}\mathbf{F}_{\mathrm{BB}}\|_F$, where $\mathbf{F}_{\mathrm{opt}}$ stands for the optimal digital fully digital beamformer, consisting of the right-singular vectors corresponding to the $N_{\mathrm{s}}$ largest singular values of the channel matrix. For the wideband OFDM system, we similarly transform the objective $\mathcal{R}=\frac{1}{K}\sum_{k=1}^K\mathcal{R}_k$ into $\frac{1}{K}\sum_{k=1}^K\|\mathbf{F}_{\mathrm{opt},k}-\mathbf{F}_\mathrm{RF}\mathbf{F}_{\mathrm{BB},k}\|_F$ to deal with the wideband scenario \cite{ofdmicc}, where $\mathbf{F}_{\mathrm{opt},k}$ is similarly defined as $\mathbf{F}_{\mathrm{opt}}$ in the narrowband scenario.

\subsection{SPEB}
From the perspective of target positioning, our primary concern is to derive the target position through the estimation of unknown parameters. The parameters to be estimated are defined as
\begin{equation} \vspace{-0.1cm} \label{eq:unknown}
    \boldsymbol{\xi}=[\boldsymbol{\xi}_0^{\mathrm{T}},\boldsymbol{\xi}_1^{\mathrm{T}},...,\boldsymbol{\xi}_L^{\mathrm{T}}]^{\mathrm{T}},
\end{equation}
with
\begin{equation} \vspace{-0.1cm}
    \boldsymbol{\xi}_l=[\theta^{\mathrm{r}}_l,\phi^{\mathrm{r}}_l,\theta^{\mathrm{t}}_l,\phi^{\mathrm{t}}_l,\tau_l,\beta_l^{\mathrm{R}},\beta_l^{\mathrm{I}}]^{\mathrm{T}},
\end{equation}
where we define $\beta_l^{\mathrm{R}}\triangleq\Re\{\beta_l\}$ and $\beta_l^{\mathrm{I}}\triangleq\Im\{\beta_l\}$. We note that the positioning process is realized by transforming the estimated parameters in \eqref{eq:unknown} into Cartesian coordinates.

The mean squared error (MSE) of an unbiased estimator $\hat{\boldsymbol{\xi}}$ is lower-bounded by CRB as \cite{kay}, \cite{iot}
\begin{equation} \vspace{-0.1cm}
    \mathbb{E}\left[(\hat{\boldsymbol{\xi}}-\boldsymbol{\xi})(\hat{\boldsymbol{\xi}}-\boldsymbol{\xi})^{\mathrm{T}}\right]\geq\mathbf{J}_{\boldsymbol{\xi}}^{-1},
\end{equation}
where $\mathbf{J}_{\boldsymbol{\xi}}\in\mathbb{R}^{7(L+1)\times7(L+1)}$ is the FIM of the unknown $\boldsymbol{\xi}$. Notably, through the derivation of sufficient statistic \cite{targetdetection}, the entries of the FIM can be written as
\begin{equation} \vspace{-0.1cm} \label{eq:FIM}     
(\mathbf{J}_{\boldsymbol{\xi}})_{i,j}=\frac{2M}{\sigma_{\mathrm{s}}^2}\Re\left\{\sum_{k=1}^K\mathrm{tr}\Big(\frac{\partial\boldsymbol{\varrho}_k}{\partial(\boldsymbol{\xi})_j}\mathbf{R}_{\mathbf{S}_k}\frac{\partial\boldsymbol{\varrho}_k^\mathrm{H}}{\partial(\boldsymbol{\xi})_i}\Big)\right\}, \end{equation} 
where 
\begin{equation} \vspace{-0.1cm}     \boldsymbol{\varrho}_k=\sqrt{E_0N_{\mathrm{t}}N_{\mathrm{r}}^{\mathrm{s}}}\beta_{l}\mathbf{b}_{l}(\theta^{\mathrm{r}}_l,\phi^{\mathrm{r}}_l)\mathbf{a}_{l}^{\mathrm{H}}(\theta^{\mathrm{t}}_l,\phi^{\mathrm{t}}_l)\mathbf{F}_{{\mathrm{RF}}}\mathbf{F}_{{\mathrm{BB},k}}, 
\end{equation}  
and $\mathbf{R}_{\mathbf{S}_k}$ is the sample coherence matrix of $\mathbf{S}_k$ approximated by 
\begin{equation} \vspace{-0.1cm}      
\mathbf{R}_{\mathbf{S}_k}\triangleq\frac{1}{M}\mathbf{S}_k\mathbf{S}_k^{\mathrm{H}} \approx\mathbb{E}[\mathbf{s}_k\mathbf{s}_k^{\mathrm{H}}]      =\frac{1}{N_{\mathrm{s}}}\mathbf{I}_{N_{\mathrm{s}}}. 
\end{equation}
Each entry of $\mathbf{J}_{\boldsymbol{\xi}}$ is calculated and listed in Appendix \ref{appen:FIM}.

\begin{figure}[t] \vspace{-0.4cm}
	\centerline{\includegraphics[width=2.5in]{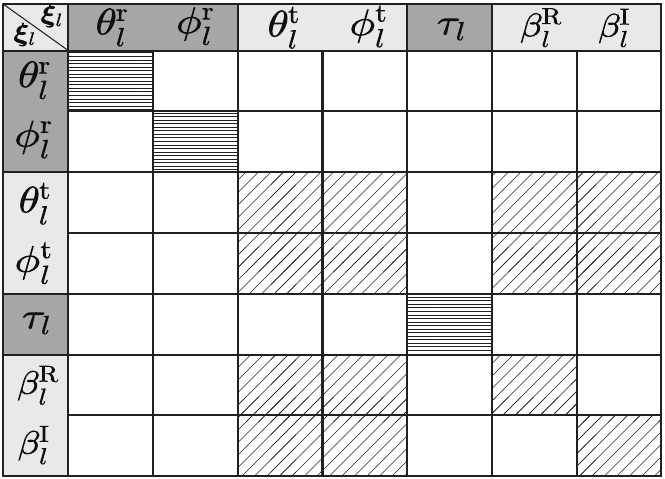}}
	\caption{Further approximation for the structure of FIM $\mathbf{J}_{{\boldsymbol{\xi}}_l}$ in \cite{henk}.}
	\label{fig:FIM}
\end{figure}

For the typical mmWave communication settings, i.e., large transmit and receive antenna arrays and large system bandwidth, multiple paths are proved orthogonal, i.e., all paths can be analyzed independently \cite{henk}, \cite{jinshi}. The FIM can thus be approximated by a block-diagonal matrix expressed as
\begin{equation} \vspace{-0.1cm} \label{eq:fimapprox}
    \mathbf{J}_{{\boldsymbol{\xi}}}\approx\begin{bmatrix}\mathbf{J}_{{\boldsymbol{\xi}}_0}&\mathbf{0}&\cdots&\mathbf{0}\\\mathbf{0}&\mathbf{J}_{{\boldsymbol{\xi}}_1}&\cdots&\mathbf{0}\\\vdots&\vdots&\ddots&\vdots\\\mathbf{0}&\mathbf{0}&\cdots&\mathbf{J}_{{\boldsymbol{\xi}}_L}\end{bmatrix},
\end{equation}
where $\mathbf{J}_{{\boldsymbol{\xi}}_l}\in\mathbb{R}^{7\times 7}$ is the $l$-th sub-matrix concerning the channel parameters of the $l$-th path, for $l=0,\ldots,L$. Taking a closer look at the submatrices, the structure of the $l$-th sub-matrix is shown in Fig. \ref{fig:FIM}, where the colored grids are non-zero elements, while the white ones denote zero elements. The detailed proof of the approximation is given in Appendix \ref{appen:approx}.

The FIM under mmWave conditions is approximated into a block-diagonal matrix in \eqref{eq:fimapprox}, where each sub-matrix exhibits a structure similar to diagonal matrices. Although the CRB of the parameters can be directly obtained by inverting the FIM $\mathbf{J}_{\boldsymbol{\xi}}$, the dimension is too high, with the unknown $\beta$ considered as a nuisance parameter. To fully exploit the structure, we first introduce the following definition:

\textit{Definition 1:} For a real parameter vector $\boldsymbol{\theta}=[\boldsymbol{\theta}_1^{\mathrm{T}},\boldsymbol{\theta}_2^{\mathrm{T}}]^{\mathrm{T}}$, with the FIM of the form
\begin{equation} \vspace{-0.1cm} \label{efimstruc}
\mathbf{J_{\boldsymbol{\theta}}}=\begin{bmatrix}\mathbf{A}&\mathbf{B}\\\mathbf{B^T}&\mathbf{A}_{\mathrm{d}}\end{bmatrix},
\end{equation}
where $\boldsymbol{\theta}\in\mathbb{R}^{U}$, $\boldsymbol{\theta}_1\in\mathbb{R}^{u}$, $\mathbf{A}\in\mathbb{R}^{u\times u}$, $\mathbf{B}\in\mathbb{R}^{u\times (U-u)}$ and $\mathbf{A}_{\mathrm{d}}\in\mathbb{R}^{(U-u)\times (U-u)}$ with $u<U$. The equivalent FIM (EFIM) of $\boldsymbol{\theta}_1$ is expressed as
\begin{equation} \vspace{-0.1cm}
    \mathbf{J}^{\mathrm{e}}_{\boldsymbol{\theta}_1}=\mathbf{A}-\mathbf{B}\mathbf{A}_{\mathrm{d}}^{-1}\mathbf{B}^{\mathrm{T}}.
\end{equation}
We note here that the EFIM is sufficient for deriving the information inequality of $\boldsymbol{\theta}_1$, and has been used to simplify the expression of CRBs \cite{moewin_bounds}, \cite{efimcrb}. Given that the paths are orthogonal, we investigate the $l$-th sub-matrix and have the following remark.

\textit{Remark 1:} The estimation of receive parameters, i.e., $\theta^{\mathrm{r}}_l$, $\phi^{\mathrm{r}}_l$ and $\tau_l$, are independent of other parameters, whereas the estimation of transmit parameters, $\theta^{\mathrm{t}}_l$ and $\phi^{\mathrm{t}}_l$, is affected by $\beta$, which harms the PEB. Specifically, when deriving the EFIM for AOD, all entries of the original FIM for AOD are reduced due to the uncertainty of $\beta$. Thus, the CRBs of AOD increase, indicating a degradation in sensing performance. Nevertheless, the EFIM for AOA and TOA isn't affected given that the off-diagonal submatrix $\mathbf{B}$ in \eqref{efimstruc} is a zero matrix. We have
\begin{equation} \vspace{-0.1cm} \label{eq:diag}
    \mathbf{J}^{\mathrm{e}}_{\boldsymbol{\eta}_l}= \mathrm{diag}(\mathbf{g}),
\end{equation}
where $\boldsymbol{\eta}_l=[\theta^{\mathrm{r}}_l,\phi^{\mathrm{r}}_l,\tau_l]^{\mathrm{T}}$ and $\mathbf{g}=[\mathbf{J}_{\boldsymbol{\xi}}(7l+1,7l+1),\mathbf{J}_{\boldsymbol{\xi}}(7l+2,7l+2),\mathbf{J}_{\boldsymbol{\xi}}(7l+5,7l+5)]^{\mathrm{T}}$.

Motivated by Remark 1, we propose a hybrid AOA/TOA positioning scheme. We note that in a bistatic ISAC system, the location of the $\text{BS}_0$ and the $\text{BS}_1$ is perfectly known, allowing the positioning based on AOA and TOA estimation. 



To better understand the relationship between the channel parameters $\boldsymbol{\eta}_l$ and the position of the $l$-th scatterer, we derive the position of the $l$-th scatterer, $\mathbf{p}_l$, w.r.t. $\theta^{\mathrm{r}}_l,\phi^{\mathrm{r}}_l$ and $\tau_l$. Specifically, denoting the distance between the $l$-th scatterer and the $\text{BS}_1$ as $d$, we have
\begin{equation} \vspace{-0.1cm} \label{coordinates}     \mathbf{p}_l=d[\sin\theta^{\mathrm{r}}_l\cos\phi^{\mathrm{r}}_l,\sin\theta^{\mathrm{r}}_l\sin\phi^{\mathrm{r}}_l,\cos\theta^{\mathrm{r}}_l]^{\mathrm{T}}, 
\end{equation} 
which shows the coordinates of the $l$-th scatterer in 3D space. Based on \eqref{coordinates} and the property of an ellipse, we write the equation with the unknown $d$ as 
\begin{equation} \vspace{-0.1cm} 
\begin{aligned}    
&c\tau_l-d=\\     
&\sqrt{d^2\sin^2\theta^{\mathrm{r}}_l\cos^2\phi^{\mathrm{r}}_l\!+\!(D\!-\!d\sin\theta^{\mathrm{r}}_l\sin\phi^{\mathrm{r}}_l)^2\!+\!d^2\cos^2\theta^{\mathrm{r}}_l}. 
\end{aligned} 
\end{equation} 
Solving the equation helps us obtain 
\begin{equation} \vspace{-0.1cm} \label{eq:coordi}    
d=\frac{c^2\tau_l^2-D^2}{2(c\tau_l-D\sin\theta^{\mathrm{r}}_l\sin\phi^{\mathrm{r}}_l)}. 
\end{equation}
Based on the insight presented by \eqref{eq:coordi}, we further give the following proposition:

\textit{Proposition 1:}  For the 3D bistatic positioning problem based on hybrid AOA/TOA estimation, we consider each path individually and derive the SPEB of the $l$-th scatterer as 
\begin{equation} \vspace{-0.1cm} \label{eq:speb}
\begin{aligned}
        \mathrm{SPEB}
        =\frac{o\times\mathrm{CRB}_{\boldsymbol{\theta}_l^{\mathrm{r}}} +p\times\mathrm{CRB}_{\boldsymbol{\phi}_l^{\mathrm{r}}} +q\times\mathrm{CRB}_{\tau_l}}{4\omega^4},
\end{aligned}
\end{equation}
where $\omega$, $o$, $p$ and $q$ are geometric coefficients defined in Appendix \ref{appen:proposition1}.

\textit{Proof:} See Appendix \ref{appen:proposition1}. $\hfill\blacksquare$

\textit{Remark 2:}
Given the closed-form expression of SPEB in \eqref{eq:speb}, we note that the only component related to the beamformer is $G=\mathbf{a}_l^{\mathrm{H}}\mathbf{F}_\mathrm{RF}\mathbf{F}_{\mathrm{BB},k}\mathbf{F}_{\mathrm{BB},k}^{\mathrm{H}}\mathbf{F}_\mathrm{RF}^{\mathrm{H}}\mathbf{a}_l$. Therefore, for the beamforming design, it is $G$ rather than the entire SPEB that should be considered.

The FIM for AOA here is a $2\times2$ matrix, given that the elevation and azimuth angle are distinct. The works in \cite{bistatic1}, which also have the same $2\times2$ FIM structure, formulate the bistatic sensing bound with a fractional form, where the denominator is a quartic function w.r.t the transmitted signal matrix. However, unlike \cite[Eq.~(18)]{bistatic1}, the denominator of the error bound under mmWave conditions appears to be quadratic as noted in Remark 2, significantly simplifying the problem. 

\subsection{Problem Formulation}
In this paper, we focus on the transmitter side and jointly design the analog beamformer $\mathbf{F}_{\mathrm{RF}}$ and the digital beamformer $\mathbf{F}_{\mathrm{BB},k}$ to achieve both communication and positioning functionalities. Most ISAC beamforming studies focus on the CRB of angles \cite{liufancrb}, \cite{wcnc}, while we combine the angles with distance and derive the PEB as a more intuitive metric for bistatic sensing. Specifically, we aim to maximize the achievable SE while satisfying a pre-defined level of PEB for the TOIs. The problem is formulated as
\begin{align}
    \min_{\mathbf{F}_{\mathrm{RF}},\{\mathbf{F}_{\mathrm{BB},k}\}_{k=1}^K}\ &\frac{1}{K}\sum_{k=1}^K\|\mathbf{F}_{\mathrm{opt},k}-\mathbf{F}_\mathrm{RF}\mathbf{F}_{\mathrm{BB},k}\|_F^2 \label{opt:1}
    \\\mathrm{s.t.~}\ \quad&\mathrm{PEB}_{i_n}\leq\Gamma,\forall i_n\in\mathcal{I}_{\mathrm{T}},\tag{\ref{opt:1}{a}} \label{opt:1a}
    \\&\|\mathbf{F}_\mathrm{RF}\mathbf{F}_{\mathrm{BB},k}\|_F^2= N_{\mathrm{s}}, \forall k,\tag{\ref{opt:1}{b}} \label{opt:1b}
    \\&|(\mathbf{F}_{\mathrm{RF}})_{i,j}|=1,\forall i,j,\tag{\ref{opt:1}{c}} \label{opt:1c}
\end{align}
where \eqref{opt:1a} represents the positioning constraint, with $\mathcal{I}_{\mathrm{T}}=\{i_1,...,i_N\}$ containing the indices of the $N$ TOIs, with $N\leq N_{\mathrm{s}}$ \cite{liuxiang}. $\mathrm{PEB}_{i_n}$ is the PEB of the $i_n$-th scatterer and $\Gamma$ denotes the pre-defined PEB threshold. By solving this problem, the waveform is properly designed to realize a reasonable communication rate with the positioning performance guaranteed.

Based on Remark 2, we claim that to satisfy the constraint \eqref{opt:1a} means to guarantee that $G$ reaches a certain value related to $\Gamma$. Therefore, we further transform the optimization problem \eqref{opt:1} into 
\begin{align} 
    \!\!\!\!\min_{\mathbf{F}_{\mathrm{RF}},\{\mathbf{F}_{\mathrm{BB},k}\}_{k=1}^K}&\frac{1}{K}\sum_{k=1}^K\|\mathbf{F}_{\mathrm{opt},k}-\mathbf{F}_\mathrm{RF}\mathbf{F}_{\mathrm{BB},k}\|_F^2 \label{opt:2}
    \\\mathrm{s.t.~}\quad\  &\!\mathbf{a}_{i_n}^{\mathrm{H}}\mathbf{F}_\mathrm{RF}\mathbf{F}_{\mathrm{BB},k}\mathbf{F}_{\mathrm{BB},k}^{\mathrm{H}}\mathbf{F}_\mathrm{RF}^{\mathrm{H}}\mathbf{a}_{i_n}\geq\kappa_n,\forall k,n,\!\tag{\ref{opt:2}{a}} \label{opt:2a}
    \\&\eqref{opt:1c}, \notag
\end{align}
where we have
\begin{equation} \vspace{-0.1cm} \label{eq:kappa}
        \!\!\kappa_n=\frac{1}{4\omega^4\gamma|\beta|^2K\Gamma^2}
        (\frac{o}{\dot{\mathbf{b}}_{\theta,i_n}^{\mathrm{H}}\dot{\mathbf{b}}_{\theta,i_n}}+\frac{p}{\dot{\mathbf{b}}_{\phi,i_n}^{\mathrm{H}}\dot{\mathbf{b}}_{\phi,i_n}}+\frac{3q}{\pi^2 B^2}).\!
\end{equation}
Here, $\dot{\mathbf{b}}_{\theta,i_n}$ and $\dot{\mathbf{b}}_{\phi,i_n}$ are defined by \eqref{eq:shortb1} and \eqref{eq:shortb2}. Notably, the power constraint is omitted here, as the objective function inherently enforces its effect \cite{alternating}. 

Problem \eqref{opt:2} is non-convex due to the non-convex set defined by constraint \eqref{opt:1c} and \eqref{opt:2a}, making jointly optimizing the analog and digital beamformers highly complicated. In the following sections, we propose algorithms to effectively solve problem \eqref{opt:2}, facilitating beamformer design for both positioning and communication.

\section{Trust-Region Based Hybrid Beamforming in Riemannian Manifold} \label{sec:mo}
In this section, we optimize the analog and digital beamformers alternatively based on the alternating minimization approach. Specifically, the shared analog beamformer $\mathbf{F}_{\mathrm{RF}}$ is designed using the trust-region method on the Riemannian manifold defined by the unit modulus constraint \eqref{opt:1c}. We first derive the Riemannian gradient and Riemannian Hessian, based on which we propose the RTR method for analog beamformer design. Meanwhile, the digital beamformer for each subcarrier, $\mathbf{F}_{\mathrm{BB},k}$, is independently optimized by approximating the difference-of-convex constraint \eqref{opt:2a} through SCA. Finally, we combine the analog and digital beamforming designs to form the overall RTR-SCA algorithm.

\subsection{RTR-based Analog Beamformer Design}
We first discuss the optimization of the analog beamformer $\mathbf{F}_\mathrm{RF}$ with fixed digital beamformers $\mathbf{F}_{\mathrm{BB},k}$. Based on the transformation $\mathrm{tr}(\mathbf{A}^{\mathrm{H}}\mathbf{P}\mathbf{A}\mathbf{C})=\mathrm{vec}(\mathbf{A})^{\mathrm{H}}(\mathbf{C}^{\mathrm{T}}\otimes\mathbf{P})\mathrm{vec}(\mathbf{A})$ and $\mathrm{vec}(\mathbf{A}\mathbf{P})=(\mathbf{P}^{\mathrm{T}}\otimes\mathbf{I})\mathrm{vec}(\mathbf{A})$, the problem can be recast as
\begin{align}
    \min_{\mathbf{x}}\ \ &\mathbf{x}^{\mathrm{H}}\mathbf{T}_1\mathbf{x}-2\Re\{\mathbf{x}^{\mathrm{H}}\mathbf{t}_1\} \label{opt:RF2}
    \\\mathrm{s.t.~}\ &\mathbf{x}^{\mathrm{H}}\boldsymbol{\Sigma}_{k,n}\mathbf{x}\geq\kappa_n,\forall k,n,\tag{\ref{opt:RF2}{a}} \label{opt:RF2a}
    \\&|\mathbf{x}_i|= 1,\forall i,\tag{\ref{opt:RF2}{b}} \label{opt:RF2b}
\end{align}
where $\mathbf{x}=\mathrm{vec}(\mathbf{F}_\mathrm{RF})$, and the matrices are defined by
\begin{subequations} \label{eq:T}
   \begin{align}
       \mathbf{T}_1&\triangleq\sum_{k=1}^K\mathbf{F}_{\mathrm{BB},k}^*\mathbf{F}_{\mathrm{BB},k}^{\mathrm{T}}\otimes\mathbf{I}_{N_{\mathrm{t}}},
       \\\mathbf{t}_1&\triangleq\sum_{k=1}^K(\mathbf{F}_{\mathrm{BB},k}^*\otimes\mathbf{I}_{N_{\mathrm{t}}})\mathbf{f}_{\mathrm{opt},k},
       \\\boldsymbol{\Sigma}_{k,n}&\triangleq\mathbf{F}_{\mathrm{BB},k}^*\mathbf{F}_{\mathrm{BB},k}^{\mathrm{T}}\otimes\mathbf{a}_{i_n}\mathbf{a}_{i_n}^{\mathrm{H}},
   \end{align}
\end{subequations}
with $\mathbf{f}_{\mathrm{opt},k}=\mathrm{vec}(\mathbf{F}_{\mathrm{opt},k})$.

The unit-modulus constraint \eqref{opt:RF2b} defines an $N_{\mathrm{t}}N_{\mathrm{RF}}$-dimensional complex circle manifold, denoted by $\mathcal{M}$ as
\begin{equation} \vspace{-0.1cm}
    \mathcal{M}=\left\{\mathbf{x}\in\mathbb{C}^{N_{\mathrm{t}}N_{\mathrm{RF}}}:|\mathbf{x}(i)|=1,\forall i\right\},
\end{equation}
which indicates that the search space of problem \eqref{opt:RF2} lies in a Riemannian sub-manifold of $\mathbb{C}^{N_{\mathrm{t}}N_{\mathrm{RF}}}$. Hence, we rewrite the problem as
\begin{align}
    \min_{\mathbf{x}\in\mathcal{M}}\ \ &\mathbf{x}^{\mathrm{H}}\mathbf{T}_1\mathbf{x}-2\Re\{\mathbf{x}^{\mathrm{H}}\mathbf{t}_1\} \label{opt:RF3}
    \\\mathrm{s.t.~}\ \;&\mathbf{x}^{\mathrm{H}}\boldsymbol{\Sigma}_{k,n}\mathbf{x}-\epsilon_{k,n}=\kappa_n,\forall k,\tag{\ref{opt:RF3}{a}} \label{opt:RF3a}
\end{align}
where $\epsilon_{k,n}=\mathrm{max}\{\mathbf{x}^{\mathrm{H}}\boldsymbol{\Sigma}_{k,n}\mathbf{x}-\kappa_n,0\}$ are non-negative slack variables. Referring to the quadratic penalty method, we combine the equality constraint with the objective function and reformulate the problem as
\begin{equation} \label{opt:mo}
    \min_{\mathbf{x}\in\mathcal{M}}\;f(\mathbf{x})=\mathbf{x}^{\mathrm{H}}\mathbf{T}_1\mathbf{x}-2\Re\{\mathbf{x}^{\mathrm{H}}\mathbf{t}_1\}+\frac{1}{2}\sum_{n=1}^N\sum_{k=1}^K\rho_{k,n} g_{k,n}^2(\mathbf{x}),
\end{equation}
where $g_{k,n}(\mathbf{x})\triangleq \mathrm{max}[0,\kappa_n-\mathbf{x}^{\mathrm{H}}\boldsymbol{\Sigma}_{k,n}\mathbf{x}]$, and $\rho_{k,n}$ are penalty coefficients dynamically adjusted to ensure that the constraint \eqref{opt:RF3a} is satisfied. 
Hence, the problem becomes finding the minimum of $f(\mathbf{x})$ on the defined complex circle manifold.

The RTR algorithm is adopted to solve problem \eqref{opt:mo} \cite{sunrui}. We first derive the crucial components, i.e., Riemannian gradient, Riemannian Hessian and Retraction. Then the RTR algorithm for analog beamforming is introduced.

The tangent space $T_{\mathbf{x}}\mathcal{M}$ can be written as
\begin{equation} \vspace{-0.1cm}
    T_{\mathbf{x}}\mathcal{M}=\left\{\mathbf{z}\in\mathbb{C}^{N_{t}N_{\mathrm{RF}}}:\Re\left\{\mathbf{z}\odot\mathbf{x}^{*}\right\}=\mathbf{0}\right\}.
\end{equation}
Among all the vectors in the tangent space $T_{\mathbf{x}}\mathcal{M}$, the one that represents the greatest decrease of the objective function is connected with the negative Riemannian gradient $\mathrm{grad}f(\mathbf{x})$, defined by
\begin{equation} \vspace{-0.1cm}
\begin{aligned}
\mathrm{grad}f(\mathbf{x})&=\mathrm{Proj}_{\mathbf{x}}\nabla f(\mathbf{x})\\&=\nabla f(\mathbf{x})-\Re\{\nabla f(\mathbf{x})\odot\mathbf{x}^*\}\odot\mathbf{x},
\end{aligned}
\end{equation}
where $\nabla f(\mathbf{x})$ is the Euclidean gradient expressed by
\begin{equation} \vspace{-0.1cm} \label{eq:gradient}
    \nabla f(\mathbf{x})=2\mathbf{T}_1\mathbf{x}-2\mathbf{t}_1
    +\sum_{n=1}^N\sum_{k=1}^K\rho_{k,n}\mathbf{q}_{k,n}(\mathbf{x}),
\end{equation}
with
\begin{equation} \vspace{-0.1cm} \label{eq:gx}
   \!\!\mathbf{q}_{k,n}(\mathbf{x})\!=\!\begin{cases}2(\mathbf{x}^{\mathrm{H}}\boldsymbol{\Sigma}_{k,n}\mathbf{x}-\kappa_n)\boldsymbol{\Sigma}_{k,n}\mathbf{x},\!\!\!&\text{if}\;\kappa_n\geq\mathbf{x}^{\mathrm{H}}\boldsymbol{\Sigma}_{k,n}\mathbf{x},\\0,&\text{otherwise}.\end{cases} 
\end{equation}
The Riemannian Hessian at point $\mathbf{x}$ along the tangent vector $\mathbf{d}$ is computed by \cite{absil2009}
\begin{equation} \vspace{-0.1cm} \label{eq:Rhessian}
    \!\!\mathrm{Hess}f(\mathbf{X})[\mathbf{d}]\!=\!\mathrm{Proj}_{\mathbf{x}}(\mathrm{D}\nabla\!f(\mathbf{x})[\mathbf{d}]-\Re\{\nabla\!f(\mathbf{x})\odot\mathbf{x}^*\}\odot\mathbf{d}).\!
\end{equation}
Here, $\mathrm{D}\nabla f(\mathbf{x})[\mathbf{d}]$ denotes the directional derivative of $\nabla f(\mathbf{x})$ along $\mathbf{d}$ and is given by
\begin{equation} \vspace{-0.1cm}
    \mathrm{D}\nabla f(\mathbf{x})[\mathbf{d}]=\nabla^2 f(\mathbf{x})\mathbf{d},
\end{equation}
where $\nabla^2 f(\mathbf{x})$ is the Euclidean Hessian matrix expressed by \eqref{eq:Ehess} shown at the bottom of next page.
\begin{figure*} [b]
\hrule
\vspace{0.5em}
\begin{equation} \label{eq:Ehess}
\nabla^2 f(\mathbf{x})=2\mathbf{T}_1+\begin{cases}2\sum_{n=1}^N\sum_{k=1}^K\rho_{k,n}(\mathbf{x}^{\mathrm{H}}\boldsymbol{\Sigma}_{k,n}\mathbf{x}-\kappa_n+\boldsymbol{\Sigma}_{k,n}\mathbf{x}\mathbf{x}^{\mathrm{H}})\boldsymbol{\Sigma}_{k,n},&\text{if}\;\kappa_n\geq\mathbf{x}^{\mathrm{H}}\boldsymbol{\Sigma}_{k,n}\mathbf{x},\\0,&\text{otherwise}.\end{cases}
\end{equation}
\end{figure*}

The notion of moving along the tangent vector $\mathbf{d}$ while remaining on the manifold is generalized by Retraction $\mathrm{R}_\mathbf{x}(\mathbf{d})$ defined by \cite{alternating}
\begin{equation} \vspace{-0.1cm} \label{eq:retraction}
    \mathrm{R}_\mathbf{x}(\mathbf{d})\triangleq\left[\frac{(\mathbf{x}+\mathbf{d})_i}{|(\mathbf{x}+\mathbf{d})_i|}\right].
\end{equation}
Based on the “lift-slove-retract” procedure in \cite{trrm}, problem \eqref{opt:mo} can be addressed via a quadratic model in the Euclidean space $T_{\mathbf{x}}\mathcal{M}$ within the trust-region framework. The trust region subproblem at the $m$-th iteration is thus expressed as
\begin{equation} \vspace{-0.1cm} \label{opt:rtr}
\begin{aligned}
    \!\min_{\mathbf{d}\in T_{\mathbf{x}_m}\mathcal{M}} w_{\mathbf{x}_m}=&f(\mathbf{x}_m)+\left\langle\mathrm{grad}f(\mathbf{x}_m),\mathbf{d}\right\rangle
    \\\!\!+\frac{1}{2}&\left\langle\mathrm{Hess}f(\mathbf{x}_m)[\mathbf{d}],\mathbf{d}\right\rangle\!, \;\mathrm{s.t.} \left\langle\mathbf{d},\mathbf{d}\right\rangle\leq \Delta_m^2,
\end{aligned}
\end{equation}
where $\Delta_m$ is the dynamic trust region radius. Whether to accept the candidate and to update the trust-region radius is based on the factor
\begin{equation} \vspace{-0.02cm} \label{eq:sk}
    s_m = \frac{f(\mathbf{x}_m) - f(\mathrm{R}_{\mathbf{x}_m}(\mathbf{d}))}{m_{\mathbf{x}_m}(0_{\mathbf{x}_m}) - m_{\mathbf{x}_m}(\mathbf{d})}.
\end{equation}
Specifically, if \(s_m\) is close to 1, the model is accurate, indicating that the update should be accepted and the radius further expanded. Conversely, if \(s_m\) is small, the update should be rejected, and the radius should be reduced. If \(s_m\) is moderate, the update can be accepted, but the radius should still be decreased. The details are presented in Algorithm \ref{alg:RTR}.

\begin{algorithm}[t]
\caption{The RTR Analog Beamforming Algorithm}\label{alg:RTR}
\renewcommand{\algorithmicrequire}{\textbf{Input:}}
\renewcommand{\algorithmicensure}{\textbf{Output:}}
\begin{algorithmic}[1]
\REQUIRE Thresholds $\epsilon_1$, $\Delta_0\in(0,\hat{\Delta})$, $\hat{s}\in[0,\frac{1}{4})$ $\delta_\rho > 1$
\STATE Initialize $\mathbf{x}$ with random phases and set $m = 1$
\REPEAT
    \STATE Compute $\mathrm{Hess}f(\mathbf{x}_m)[\mathbf{d}]$ based on \eqref{eq:Rhessian}
    \STATE Obtain $\mathbf{d}$ by solving \eqref{opt:rtr} and compute $s_m$ via \eqref{eq:sk}
    \STATE $\Delta_{m+1}=\begin{cases}
        \frac{1}{4}\Delta_m,&\text{if}\;s_m<\frac{1}{4},\\\mathrm{min}\{2\Delta_m,\hat{\Delta}\}&\text{if}\;s_m>\frac{3}{4}$ \textbf{and} $\|\mathbf{d}\|=\Delta_m,\\\Delta_m,&\text{otherwise},
    \end{cases}$
    \STATE $\mathbf{x}_{m+1}=\begin{cases}
        \mathrm{R}_{\mathbf{x}_m}(\mathbf{d}),&\text{if}\;s_m>\hat{s},\\\mathbf{x}_{n},&\text{otherwise},
    \end{cases}$
    \STATE $m \leftarrow m+1$
\UNTIL convergence
\FOR{$n\leq N$} 
    \IF{$\exists\;k$ such that $g_{k,n}(\mathbf{x}_{m+1}) \geq \epsilon_1$}
        \STATE Update $\rho_{k,n} = \delta_\rho \rho_{k,n}, \forall k$, and repeat from Step 2
    \ENDIF
\ENDFOR
\STATE Reshape $\mathbf{x}_m$ into $\mathbf{F}_{\mathrm{RF}}$
\ENSURE $\mathbf{F}_{\mathrm{RF}}$
\end{algorithmic}
\end{algorithm}


\subsection{SCA-based Digital Beamformer Design}
According to principles of alternating minimization, $\mathbf{F}_{\mathrm{BB},k}$ is optimized with a given analog beamformer $\mathbf{F}_{\mathrm{RF}}$. Thus, problem \eqref{opt:2} is simplified into
\begin{align} 
    \min_{\{\mathbf{F}_{\mathrm{BB},k}\}_{k=1}^K}\;&\frac{1}{K}\sum_{k=1}^K\|\mathbf{F}_{\mathrm{opt},k}-\mathbf{F}_\mathrm{RF}\mathbf{F}_{\mathrm{BB},k}\|_F^2 \label{opt:3}
    \\\mathrm{s.t.~}\ \;&\mathbf{a}_{i_n}^{\mathrm{H}}\mathbf{F}_\mathrm{RF}\mathbf{F}_{\mathrm{BB},k}\mathbf{F}_{\mathrm{BB},k}^{\mathrm{H}}\mathbf{F}_\mathrm{RF}^{\mathrm{H}}\mathbf{a}_{i_n}\geq\kappa_n,\forall k,n.\tag{\ref{opt:3}{a}} \label{opt:3a}
\end{align}
Upon closer examination of problem \eqref{opt:3}, we observe that the optimization of $\mathbf{F}_{\mathrm{BB},k}$ for each subcarrier is independent and can be performed in parallel. We express the optimization of $\mathbf{f}_{\mathrm{BB},k}=\mathrm{vec}(\mathbf{F}_{\mathrm{BB},k})$ as
\begin{align} 
    \min_{\mathbf{f}_{\mathrm{BB},k}}\ \  &\mathbf{f}_{\mathrm{BB},k}^{\mathrm{H}}\mathbf{R}_1\mathbf{f}_{\mathrm{BB},k}-2\Re\{\mathbf{f}_{\mathrm{BB},k}^{\mathrm{H}}\mathbf{R}_2\mathbf{f}_{\mathrm{opt}}\} \label{opt:k1}
    \\\mathrm{s.t.~}\ \;&\mathbf{f}_{\mathrm{BB},k}^{\mathrm{H}}\mathbf{R}_3\mathbf{f}_{\mathrm{BB},k}\geq\kappa_n.\tag{\ref{opt:k1}{a}} \label{opt:k1a}
\end{align}
Here, the matrices are respectively defined by $\mathbf{R}_1\triangleq\mathbf{I}_{N_{\mathrm{s}}}\otimes\mathbf{F}_\mathrm{RF}^{\mathrm{H}}\mathbf{F}_\mathrm{RF}$, $\mathbf{R}_2\triangleq\mathbf{I}_{N_{\mathrm{s}}}\otimes\mathbf{F}_\mathrm{RF}^{\mathrm{H}}$, and $\mathbf{R}_3\triangleq\mathbf{I}_{N_{\mathrm{s}}}\otimes\mathbf{F}_\mathrm{RF}^{\mathrm{H}}\mathbf{a}_{i_n}\mathbf{a}_{i_n}^{\mathrm{H}}\mathbf{F}_\mathrm{RF}$. 

The objective function is convex, but the constraint \eqref{opt:k1a} defines a non-convex set. Since the left-hand side of \eqref{opt:k1a} is lower-bounded by its first-order Taylor expansion, we have
\begin{equation} \vspace{-0.1cm}
    \mathbf{f}_{\mathrm{BB},k}^{\mathrm{H}}\mathbf{R}_3\mathbf{f}_{\mathrm{BB},k}\!\geq\!2\Re\{\mathbf{f}_{\mathrm{BB},k}^{\mathrm{H}}\mathbf{R}_3\mathbf{f}_{\mathrm{BB},k}^{(i)}\}-\mathbf{f}_{\mathrm{BB},k}^{(i)\,\mathrm{H}}\mathbf{R}_3\mathbf{f}_{\mathrm{BB},k}^{(i)},
\end{equation}
where $\mathbf{f}_{\mathrm{BB},k}^{(i)}$ is a given point in the $i$-th iteration. Motivated by the idea of SCA, we approximate the non-convex quadratic constraint \eqref{opt:k1a} as
\begin{equation} \vspace{-0.1cm} \label{eq:sca}
    2\Re\{\mathbf{f}_{\mathrm{BB},k}^{\mathrm{H}}\mathbf{R}_3\mathbf{f}_{\mathrm{BB},k}^{(i)}\}-\mathbf{f}_{\mathrm{BB},k}^{(i)\,\mathrm{H}}\mathbf{R}_3\mathbf{f}_{\mathrm{BB},k}^{(i)}\geq\kappa_n.
\end{equation}
Thus, the problem for the $i$-th iteration of the SCA-based algorithm is formulated as
\begin{align} 
    \min_{\mathbf{f}_{\mathrm{BB},k}}\ \  &\mathbf{f}_{\mathrm{BB},k}^{\mathrm{H}}\mathbf{R}_1\mathbf{f}_{\mathrm{BB},k}-2\Re\{\mathbf{f}_{\mathrm{BB},k}^{\mathrm{H}}\mathbf{R}_2\mathbf{f}_{\mathrm{opt}}\} \label{opt:k3}
    \\\mathrm{s.t.~}\ \;&\eqref{eq:sca}. \notag
\end{align}
Notably, problem \eqref{opt:k3} is now convex and can be solved via numerical tools such as CVX \cite{cvx}.

\subsection{Overall Algorithm}
In the subsections above, we clarify the RTR-based analog beamformer design algorithm and the SCA-based digital beamformer design algorithm, which are combined to form the RTR-SCA algorithm, as shown in Algorithm \ref{alg:RTR-SCA}.
\begin{algorithm}[t]
\caption{The RTR-SCA Beamforming Algorithm}\label{alg:RTR-SCA}
\renewcommand{\algorithmicrequire}{\textbf{Input:}}
\renewcommand{\algorithmicensure}{\textbf{Output:}}
\begin{algorithmic}[1]
\REQUIRE Thresholds $\epsilon_2$
\REPEAT
\STATE Initialize $\mathbf{f}_{\mathrm{BB},k}^{(1)}$, set $m = 1$ and $O^{(0)}=0$.
\REPEAT 
    \STATE Obtain $\mathbf{f}_{\mathrm{BB},k}$ by solving problem \eqref{opt:k1}
    \STATE Compute the objective function $O^{(m)}(\mathbf{f}_{\mathrm{BB},k})$ in \eqref{opt:k1}
    \STATE $\mathbf{f}_{\mathrm{BB},k}^{(m+1)}\leftarrow\mathbf{f}_{\mathrm{BB},k}$
    \STATE $m \leftarrow m + 1$
\UNTIL $|O^{(m)}(\mathbf{f}_{\mathrm{BB},k})-O^{(m-1)}(\mathbf{f}_{\mathrm{BB},k})| \leq \epsilon_2$
\STATE Reshape $\mathbf{f}_{\mathrm{BB},k}$ as $\hat{\mathbf{F}}_{\mathrm{BB},k},\forall k$
\STATE Update $\mathbf{F}_{\mathrm{RF}}$ via Algorithm \ref{alg:RTR}
\UNTIL the Euclidean distance in problem \eqref{opt:2} converges
\STATE $\mathbf{F}_{\mathrm{BB},k}=\sqrt{N_{\mathrm{s}}}\frac{\hat{\mathbf{F}}_{\mathrm{BB},k}}{\|\mathbf{F}_{\mathrm{RF}}\hat{\mathbf{F}}_{\mathrm{BB},k}\|_F}, 1\leq k\leq K$
\ENSURE $\mathbf{F}_{\mathrm{RF}}$, $\{\mathbf{F}_{\mathrm{BB},k}\}_{k=1}^K$
\end{algorithmic}
\end{algorithm}

The complexity of each inner iteration for Algorithm \ref{alg:RTR-SCA} is mainly due to the Riemannian Hessian computation in \eqref{eq:Rhessian}, which has a complexity of $\mathcal{O}(KN_{\mathrm{RF}}^2N_{\mathrm{t}}^2)$. Thus, assuming the algorithm undergoes $N_{\mathrm{o}}$ outer iterations and $N_{\mathrm{i}}$ inner iterations, the total computational complexity is $\mathcal{O}(N_{\mathrm{o}}N_{\mathrm{i}}KN_{\mathrm{RF}}^2N_{\mathrm{t}}^2)$.

\section{Low-Complexity Hybrid Beamforming for Positioning and Communication} \label{sec:omp}
While the RTR-SCA algorithm effectively addresses the unit-modulus constraint using Riemannian manifold techniques, its computational complexity is relatively high in mmWave MIMO-OFDM systems. For practical implementation, this complexity must be reduced without significantly degrading positioning and communication performance. In this section, we refine the classical OMP-based algorithm from \cite{spatially} for positioning capabilities and propose a low-complexity algorithm that simultaneously realizes both communication and positioning functionalities. This approach significantly reduces computational complexity, with only a modest impact on communication performance.

\subsection{Partition of the Beamforming Matrix}
The problem without PEB constraint \eqref{opt:2a} is written as
\begin{align} 
    \min_{\mathbf{F}_{\mathrm{RF}},\{\mathbf{F}_{\mathrm{BB},k}\}_{k=1}^K}\ &\frac{1}{K}\sum_{k=1}^K\|\mathbf{F}_{\mathrm{opt},k}-\mathbf{F}_\mathrm{RF}\mathbf{F}_{\mathrm{BB},k}\|_F^2
    \\\mathrm{s.t.~}\quad\quad&\eqref{opt:1c}. \notag
\end{align}
 and is solved by applying the classical OMP-based method to select $N_{\mathrm{RF}}$ columns from the dictionary matrix $\mathbf{A}_{\mathrm{d}}$ given by
\begin{equation} \vspace{-0.1cm}
\mathbf{A}_{\mathrm{d}}\!=\!\sqrt{N_{\mathrm{t}}}\!\left[\mathbf{a}_{1\!,1}(\theta_{1\!,1}^\mathrm{t},\!\phi_{1\!,1}^\mathrm{t}),...,\mathbf{a}_{N_{\mathrm{cl}},N_{\mathrm{ray}}}(\theta_{N_{\mathrm{cl}},N_{\mathrm{ray}}}^\mathrm{t},\!\phi_{N_{\mathrm{cl}}, N_{\mathrm{ray}}}^\mathrm{t})\right],
\end{equation}
where $\mathbf{A}_{\mathrm{d}}$ is composed of all steering vectors from the transmit side of the communication channel. The OMP-based algorithm leverages the orthogonality between the steering vectors \cite{spatially}.

The PEB constraint \eqref{opt:2a} requires the beamformer to ensure that its projection onto the direction of $\mathbf{a}_{i_n}$ meets a specified threshold. Considering that under large antenna arrays, where the beams become extremely narrow, we have \cite{jinshi}
\begin{equation} \vspace{-0.1cm} \label{eq:ortho}
    \mathbf{a}_{i_n}^{\mathrm{H}}\mathbf{a}_{i_n}\gg\mathbf{a}_{i_n}^{\mathrm{H}}\mathbf{a}_{ij}(\theta_{ij}^\mathrm{t},\phi_{ij}^\mathrm{t}),\forall i,j,n,
\end{equation}
which reveals the orthogonality between $\mathbf{a}_{i_n}$ and the steering vectors in $\mathbf{A}_{\mathrm{d}}$. As a result, fixing columns of $\mathbf{F}_{\mathrm{RF}}$ as $\sqrt{N_{\mathrm{t}}}\mathbf{a}_{i_n}$ becomes feasible and necessary since the vectors selected from $\mathbf{A}_{\mathrm{d}}$ do not contribute to positioning performance. To simultaneously guarantee the positioning and communication performance, we partition the beamforming matrix as
\begin{subequations} \label{eq:partition}
\begin{align}
    &\mathbf{F}_{\mathrm{RF}}=[\mathbf{A}_{\mathrm{s}},\mathbf{F}_{\mathrm{RF}}^{\mathrm{ex}}], \\ &\mathbf{F}_{\mathrm{BB},k}=[\mathbf{F}_{\mathrm{s},k},\mathbf{F}_{\mathrm{BB},k}^{\mathrm{ex}\,\mathrm{T}}]^{\mathrm{T}}, 1\leq k\leq K.
\end{align}
\end{subequations}
Here, $\mathbf{A}_{\mathrm{s}}=\sqrt{N_{\mathrm{t}}}[\mathbf{a}_{i_1},...,\mathbf{a}_{i_N}]$ is composed of the steering vectors corresponding to all TOIs. $\mathbf{F}_{\mathrm{s},k}\in\mathbb{C}^{N_{\mathrm{s}}\times N}$ is the part of the digital beamformer corresponding to $\mathbf{A}_{\mathrm{s}}$, which is designed for positioning. $\mathbf{F}_{\mathrm{RF}}^{\mathrm{ex}}\in\mathbb{C}^{N_{\mathrm{t}}\times (N_{\mathrm{RF}}-N)}$ represents the $N_{\mathrm{RF}}-N$ steering vectors selected from $\mathbf{A}_{\mathrm{d}}$, while $\mathbf{F}_{\mathrm{BB},k}^{\mathrm{ex}}\in\mathbb{C}^{(N_{\mathrm{RF}}-N)\times N_{\mathrm{s}}}$ corresponds to $\mathbf{F}_{\mathrm{RF}}^{\mathrm{ex}}$ and is designed for communication. Notably, to maintain sufficient degrees of freedom for communication, it is recommended that the number of RF chains satisfies $N_{\mathrm{RF}}\geq N_\mathrm{s}+N$. The beamformer of the $k$-th subcarrier can therefore be expressed by
\begin{equation} \vspace{-0.1cm}
\mathbf{F}_k=\mathbf{A}_{\mathrm{s}}\mathbf{F}_{\mathrm{s},k}^{\mathrm{T}}+\mathbf{F}_{\mathrm{RF}}^{\mathrm{ex}}\mathbf{F}_{\mathrm{BB},k}^{\mathrm{ex}},
\end{equation}
where the first term represents the positioning side and the second denotes the communication side. The design algorithm will be presented below, along with an illustration of the independence between the two sides introduced by the partition.

\subsection{Design of the Positioning Side}
Upon closer examination of the PEB constraint, we rewrite the left-hand side of \eqref{opt:2a} as
\begin{equation} \vspace{-0.1cm}
    \begin{aligned}
        \mathbf{a}_{i_n}^{\mathrm{H}}\mathbf{F}_\mathrm{RF}&\mathbf{F}_{\mathrm{BB},k}\mathbf{F}_{\mathrm{BB},k}^{\mathrm{H}}\mathbf{F}_\mathrm{RF}^{\mathrm{H}}\mathbf{a}_{i_n}
        \\&=\mathbf{a}_{i_n}^{\mathrm{H}}(\mathbf{A}_{\mathrm{s}}\mathbf{F}_{\mathrm{s},k}^{\mathrm{T}}+\mathbf{F}_{\mathrm{RF}}^{\mathrm{ex}}\mathbf{F}_{\mathrm{BB},k}^{\mathrm{ex}})\\&\quad\quad\times(\mathbf{A}_{\mathrm{s}}\mathbf{F}_{\mathrm{s},k}^{\mathrm{T}}+\mathbf{F}_{\mathrm{RF}}^{\mathrm{ex}}\mathbf{F}_{\mathrm{BB},k}^{\mathrm{ex}})^{\mathrm{H}}\mathbf{a}_{i_n}
        \\&\overset{(a)}{\approx}N_{\mathrm{t}}\mathbf{F}_{\mathrm{s},k}^{\mathrm{T}}(:,n)\mathbf{F}_{\mathrm{s},k}^{*}(:,n).
    \end{aligned}
\end{equation}
Here, approximation (a) follows from the orthogonality presented by \eqref{eq:ortho}, indicating that the beamformer designed for communication, $\mathbf{F}_{\mathrm{RF}}^{\mathrm{ex}}$ and $\mathbf{F}_{\mathrm{BB}}^{\mathrm{ex}}$, has only a slight impact on positioning, which can be ignored.

Therefore, the PEB constraint can be transformed into
\begin{equation} \vspace{-0.1cm}
    \|\mathbf{F}_{\mathrm{s},k}(:,n)\|^2\geq v_n,\forall
    k,n,
\end{equation}
where $v_n=\frac{\kappa_n}{N_{\mathrm{t}}}$. 
We then allocate power by setting $\mathbf{F}_{\mathrm{s},k}$ as
\begin{equation} \vspace{-0.1cm}
    \mathbf{F}_{\mathrm{s},k}(:,n)=\sqrt{v_n}\frac{\boldsymbol{\psi}}{\|\boldsymbol{\psi}\|},
\end{equation}
where $\boldsymbol{\psi}= \mathbf{a}_{i_n}^{\mathrm{H}}\mathbf{F}_{\mathrm{opt},k}$ is the projection of $\mathbf{F}_{\mathrm{opt},k}$ onto the direction of $\mathbf{a}_{i_n}$.

\subsection{Design of the Communication Side}
The design on the positioning side ensures that the positioning requirements are met. For the communication design, we define the communication design matrix as
\begin{equation} \vspace{-0.1cm}
    \mathbf{F}_{\mathrm{opt},k}^{\mathrm{ex}}=\mathbf{F}_{\mathrm{opt},k}-\mathbf{A}_{\mathrm{s}}\mathbf{F}_{\mathrm{s},k}^{\mathrm{T}}.
\end{equation}
To approximate $\mathbf{F}_{\mathrm{opt},k}^{\mathrm{ex}}$, the steering vector along which $\mathbf{F}_{\mathrm{opt},k}^{\mathrm{ex}}$ has the maximum projection is selected as the first column of $\mathbf{F}_{\mathrm{RF}}^{\mathrm{ex}}$. After identifying the dominant vector, the least squares solution for $\hat{\mathbf{F}}_{\mathrm{BB},k}^{\mathrm{ex}}$ is computed. The algorithms then proceed to find the dominant vector of the residual matrix defined by
\begin{equation} \vspace{-0.1cm}
    \mathbf{F}_{\mathrm{res},k}=\frac{\mathbf{F}_{\mathrm{opt}}^{\mathrm{ex}}-\mathbf{F}_{\mathrm{RF}}^{\mathrm{ex}}\hat{\mathbf{F}}_{\mathrm{BB},k}^{\mathrm{ex}}}{\|\mathbf{F}_{\mathrm{opt}}^{\mathrm{ex}}-\mathbf{F}_{\mathrm{RF}}^{\mathrm{ex}}\hat{\mathbf{F}}_{\mathrm{BB},k}^{\mathrm{ex}}\|_F}, 1\leq k\leq K.
\end{equation}
The process is repeated until $N_{\mathrm{RF}}-N$ steering vectors are found, and $\hat{\mathbf{F}}_{\mathrm{BB},k}^{\mathrm{ex}}$ is then normalized as
\begin{equation} \vspace{-0.1cm}
    \mathbf{F}_{\mathrm{BB},k}^{\mathrm{ex}}=\|\mathbf{F}_{\mathrm{opt},k}^{\mathrm{ex}}\|_F\frac{\hat{\mathbf{F}}_{\mathrm{BB},k}^{\mathrm{ex}}}{\|\mathbf{F}_{\mathrm{RF}}^{\mathrm{ex}}\hat{\mathbf{F}}_{\mathrm{BB},k}^{\mathrm{ex}}\|_F}, 1\leq k\leq K.
\end{equation}
With the communication beamformer designed according to the aforementioned procedures, the entire beamformer is combined as shown in \eqref{eq:partition}, completing the design algorithm.

Notably, although the normalization step confirms that initially ignoring the power constraint is reasonable \cite{alternating}, combining the communication design beamformer with $\mathbf{A}_{\mathrm{s}}$ and $\mathbf{F}_{\mathrm{s},k}$ may invalidate the power constraint. To demonstrate that the power constraint remains satisfied after the combination, we present the following lemma.

\textit{Lemma 2:} If the Euclidean distance before normalization satisfies $\|\mathbf{F}_{\mathrm{opt}}^{\mathrm{ex}}-\mathbf{F}_{\mathrm{RF}}^{\mathrm{ex}}\hat{\mathbf{F}}_{\mathrm{BB},k}^{\mathrm{ex}}\|_F\leq\delta$, then after normalization and combining in \eqref{eq:partition}, the final beamformer satisfies $|\|\mathbf{F}_{\mathrm{RF}}\mathbf{F}_{\mathrm{BB},k}\|_F-\sqrt{N_{\mathrm{s}}}|\leq2\delta$.

\textit{Proof:} See Appendix \ref{appen:norm}. $\hfill\blacksquare$

\subsection{Overall Algorithm}
In the subsections above, we illustrate that the partition in \eqref{eq:partition} is effective in a way that the positioning and the communication part are approximately independent of each other. Combining the positioning and communication design procedures, the PC-OMP algorithm is given in detail by Algorithm \ref{alg:pc-omp}.
\begin{algorithm}[t]
\caption{The PC-OMP Beamforming Algorithm}\label{alg:pc-omp}
\renewcommand{\algorithmicrequire}{\textbf{Input:}}
\renewcommand{\algorithmicensure}{\textbf{Output:}}
\begin{algorithmic}[1]
\REQUIRE $\{\mathbf{F}_{\mathrm{opt},k}\}_{k=1}^K$, $\mathbf{A}_{\mathrm{s}}$, $\mathbf{A}_{\mathrm{d}}$, $\{\kappa_n\}_{N=1}^N$ 
\STATE$ \mathbf{F}_{\mathrm{RF}}= \mathbf{A}_{\mathrm{s}}$ and $ v_n=\frac{\kappa_n}{N_{\mathrm{t}}},\forall n$
\FOR{$k\leq K$}
\FOR{$n \leq N$}
\STATE$ \boldsymbol{\psi}= \mathbf{F}_{\mathrm{RF}}^{\mathrm{H}}(:,n)\mathbf{F}_{\mathrm{opt},k}$
    \STATE $\mathbf{F}_{\mathrm{s},k}(:,n)=\sqrt{v_n}\frac{\boldsymbol{\psi}}{\|\boldsymbol{\psi}\|}$
    \ENDFOR
\STATE $\mathbf{F}_{\mathrm{opt},k}^{\mathrm{ex}}=\mathbf{F}_{\mathrm{opt},k}-\mathbf{A}_{\mathrm{s}}\mathbf{F}_{\mathrm{s},k}^{\mathrm{T}}$
\ENDFOR
\STATE $\mathbf{F}_{\mathrm{res},k}=\mathbf{F}_{\mathrm{opt},k}^{\mathrm{ex}},1\leq k \leq K$
\FOR{$i\leq N_{\mathrm{RF}}-N$} 
    \STATE $\boldsymbol{\Pi}_k=\mathbf{A}_{\mathrm{d}}^{\mathrm{H}}\mathbf{F}_{\mathrm{res},k}, 1\leq k\leq K$
    \STATE $q=\arg\max_{m=1,...,N_{\mathrm{cl}}N_{\mathrm{ray}}}\left(\sum_{k=1}^{K}\boldsymbol{\Pi}_k\boldsymbol{\Pi}_k^{H}\right)_{m,m}$
    \STATE $\mathbf{F}_{\mathrm{RF}}^{\mathrm{ex}}=[\mathbf{F}_{\mathrm{RF}}^{\mathrm{ex}},\sqrt{N_{\mathrm{t}}}\mathbf{A}_{\mathrm{d}}(:,q)]$
    \STATE $\hat{\mathbf{F}}_{\mathrm{BB},k}^{\mathrm{ex}}=(\mathbf{F}_{\mathrm{RF}}^{\mathrm{ex}\,\mathrm{H}}\mathbf{F}_{\mathrm{RF}}^{\mathrm{ex}})^{-1}\mathbf{F}_{\mathrm{RF}}^{\mathrm{ex}\,\mathrm{H}}\mathbf{F}_{\mathrm{opt}}^{\mathrm{ex}}$
    \STATE $\mathbf{F}_{\mathrm{res},k}=\frac{\mathbf{F}_{\mathrm{opt}}^{\mathrm{ex}}-\mathbf{F}_{\mathrm{RF}}^{\mathrm{ex}}\hat{\mathbf{F}}_{\mathrm{BB},k}^{\mathrm{ex}}}{\|\mathbf{F}_{\mathrm{opt}}^{\mathrm{ex}}-\mathbf{F}_{\mathrm{RF}}^{\mathrm{ex}}\hat{\mathbf{F}}_{\mathrm{BB},k}^{\mathrm{ex}}\|_F}, 1\leq k\leq K$
\ENDFOR
\STATE $\mathbf{F}_{\mathrm{BB},k}^{\mathrm{ex}}=\|\mathbf{F}_{\mathrm{opt},k}^{\mathrm{ex}}\|_F\frac{\hat{\mathbf{F}}_{\mathrm{BB},k}^{\mathrm{ex}}}{\|\mathbf{F}_{\mathrm{RF}}^{\mathrm{ex}}\hat{\mathbf{F}}_{\mathrm{BB},k}^{\mathrm{ex}}\|_F}, 1\leq k\leq K$
\STATE $\mathbf{F}_{\mathrm{RF}}=[\mathbf{A}_{\mathrm{s}},\mathbf{F}_{\mathrm{RF}}^{\mathrm{ex}}]$
\STATE $\mathbf{F}_{\mathrm{BB},k}=[\mathbf{F}_{\mathrm{s},k},\mathbf{F}_{\mathrm{BB},k}^{\mathrm{ex}\,\mathrm{T}}]^{\mathrm{T}}, 1\leq k\leq K$
\ENSURE $\mathbf{F}_{\mathrm{RF}}$, $\{\mathbf{F}_{\mathrm{BB},k}\}_{k=1}^K$    
\end{algorithmic}
\label{alg1}
\end{algorithm}

The complexity of the positioning beamformer design is $\mathcal{O}\left[KN_{\mathrm{t}}(N_{\mathrm{s}}+N_{\mathrm{RF}})\right]$, dominated by step 3 and 8 in Algorithm \ref{alg:pc-omp}. The complexity of the communication beamformer design, contributed mainly by step 11 and 14, is $\mathcal{O}\left[KN_{\mathrm{t}}N_{\mathrm{s}}N_{\mathrm{RF}}(N_{\mathrm{RF}}+N_{\mathrm{cl}}N_{\mathrm{ray}})\right]$. Thus, the total computational complexity can be approximated by the latter component $\mathcal{O}\left[KN_{\mathrm{t}}N_{\mathrm{s}}N_{\mathrm{RF}}(N_{\mathrm{RF}}+N_{\mathrm{cl}}N_{\mathrm{ray}})\right]$, which is much lower than that of the RTR-SCA algorithm.

\section{Simulation Results} \label{sec:simulation}
In this section, numerical results are presented to illustrate our findings. The base station $\text{BS}_0$, equipped with $N_{\mathrm{t}} = 100$ transmit antennas, sends signals to a communication receiver with $N_{\mathrm{r}}^{\mathrm{c}} = 100$ antennas. Additionally, $\text{BS}_1$, which has $N_{\mathrm{r}}^{\mathrm{s}} = 100$ receive antennas, captures signals scattered from the environment and performs positioning for the TOIs. $\text{BS}_0$ and $\text{BS}_1$ are separated by a distance of $D = 200$ m, with their $10 \times 10$ USPAs positioned at a height of $h = 10$ m. The array geometry in 3D space is modeled such that the USPA of $\text{BS}_1$ is centered at $(0, D, 0)$ and directly faces the USPA of $\text{BS}_0$, which has zero orientation. Similarly, the USPA of $\text{BS}_0$ is centered at $(0, 0, 0)$ and faces $\text{BS}_1$. We set $L=4$ for the sensing channel, with the four scatterers located at the coordinates $(60, 100, -10)$, $(70, 50, 0)$, $(10, 0, 20)$ and $(-60, 150, 30)$, respectively. For $N=1$, the single TOI is the scatterer at $(60, 100, -10)$, while for $N=2$, the TOIs are located at $(60, 100, -10)$ and $(-60, 150, 30)$. The OFDM system uses $K = 128$ subcarriers operating at $f = 28$ GHz, with the subcarrier spacing $\Delta F = 240$ kHz. For positioning, we utilize $M = 30$ symbols, each with a transmit power of $E_0/T_{\mathrm{S}} = 37$ dBm. The noise power is set to $\sigma_{\mathrm{s}}^2 = -83$ dBm.

\subsection{PEB Analysis}
\begin{figure}[t] \vspace{-0.4cm} 
	\centerline{\includegraphics[width=3.4in]{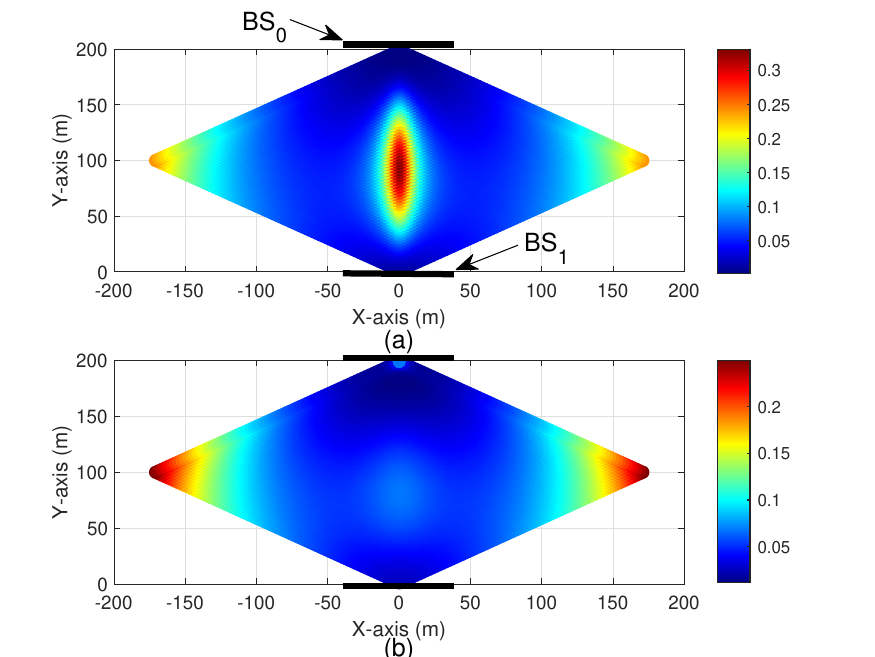}}
	\caption{PEB for targets at (a) $z=-10$. (b) $z=30$.}
	\label{fig:peb}
\end{figure}

We start by investigating the PEB for the targets within a flat 120 sector of a sectorized cell with a radius of $D$ meters. To gain fundamental understandings, we adopt a simple analog beamforming here, which can be written as
\begin{equation} \vspace{-0.1cm}
    \mathbf{F}=\sqrt{N_{\mathrm{s}}}\mathbf{a}(\theta_i,\phi_i),
\end{equation}
where $\theta_i$ and $\phi_i$ are the AOD of the $i$-th point target in the sector. Fig. \ref{fig:peb} shows the PEB as a function of the target location under the hybrid AOA/TOA positioning scheme, with $N_{\mathrm{t}}=N_{\mathrm{r}}^{\mathrm{c}} = 100$. The sectors are set at $z=-10$ m and $z=30$ m, representing ground and aerial targets. Generally, PEB worsens with increased target path length. From \eqref{eq:o} and \eqref{eq:p}, we observe that PEB increases with $c^2\tau^2-D^2$, which aligns with the numerical results. However, positioning performance deteriorates when the target is close to the line connecting the two arrays, as shown by Fig. \ref{fig:peb}(a). Numerically, this degradation is due to $\omega$ in the denominator of the SPEB expression \eqref{eq:speb}. As the target gets closer to the line, $\omega$ approaches 0, causing PEB to increase. This outcome is intuitive, as when delay indicates that the target is positioned along the line connecting the two array planes, positioning the point target by angles becomes challenging. No such degradation is seen in the sector at $z=30$ far from the line. 



\begin{figure}[t] \vspace{-0.4cm} 
	\centerline{\includegraphics[width=3.4in]{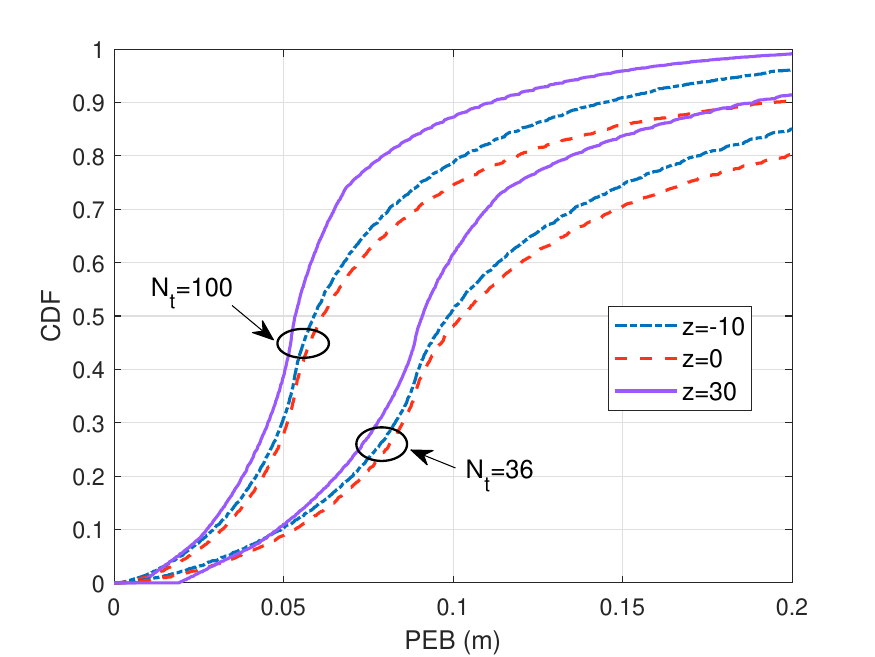}}
	\caption{CDF of the PEB for targets located at $z=-10$, $z=0$ and $z=30$ with $N_{\mathrm{r}}^{\mathrm{s}}=100$.}
	\label{fig:cdf}
\end{figure}

Next, we collect the PEB values of all target points within the sector and compute the cumulative distribution function (CDF). Fig. \ref{fig:cdf} presents the CDF for targets located in three sectors: \( z = -10 \), \( z = 0 \), and \( z = 30 \), representing ground targets, targets at the same height as the BS, and aerial targets, respectively. For \( N_{\mathrm{t}} = 36 \), targets at \( z = 0 \), which are closest to the line connecting the two arrays, exhibit the poorest CDF performance, with only 74.5\% achieving a PEB below 0.1 m. As the target height deviates further from \( z = 0 \), the CDF curve shows a marked improvement. Furthermore, considering the high hardware cost and computational complexity of large antenna arrays, we reduce the transmit array size by setting \( N_{\mathrm{t}} = 36 \). This reduction significantly impacts positioning performance, with only about half of the targets achieving a PEB below 0.1 m for all sectors, which is attributed to the influence of \( N_{\mathrm{t}} \) on the SNR factor \( \gamma \).

\subsection{SE Analysis}
\begin{figure}[t] \vspace{-0.4cm} 
	\centerline{\includegraphics[width=3.4in]{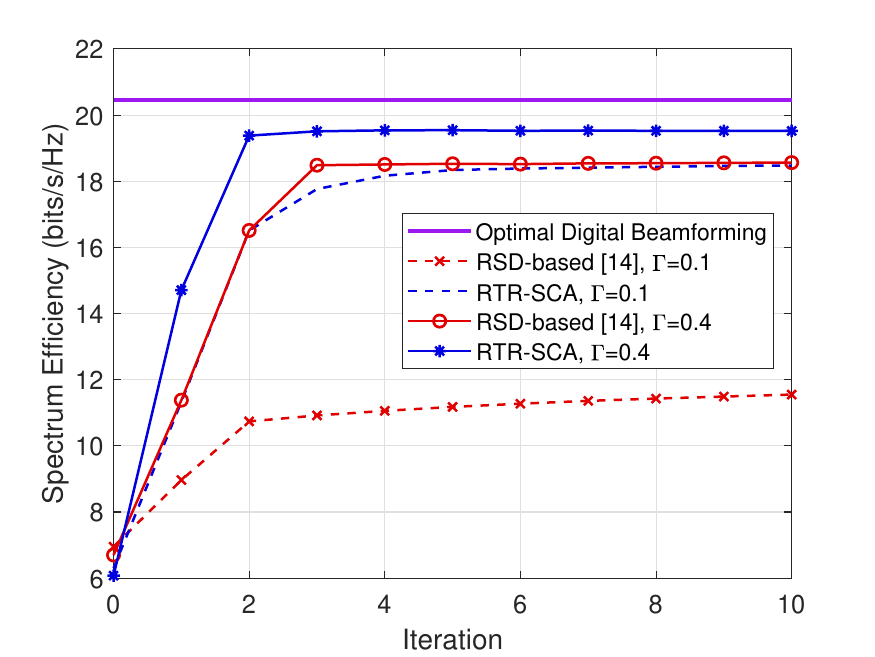}}
	\caption{Convergence of the RTR-SCA and RSD-based algorithm in \cite{partially} when $N=1$, $N_{\mathrm{s}}=N_{\mathrm{RF}}=2$, and $\text{SNR}_\mathrm{c}=0$ dB.}
	\label{fig:convergence}
\end{figure}

We start with analyzing the convergence of the proposed RTR-SCA algorithm. For comparison, we also apply the algorithm based on Riemannian Steepest Descent (RSD) from \cite{partially} to solve problem \eqref{opt:2}. Fig. \ref{fig:convergence} shows the convergence behavior of the proposed algorithm compared to the RSD-based algorithm from \cite{partially} when $N=1$, \( N_{\mathrm{s}} = 2 \) and \( \text{SNR}_{\mathrm{c}} = 0 \) dB, with the x-axis denoting the number of outer iterations. As shown, when the PEB threshold is set to \( \Gamma = 0.1 \) m, representing a strong constraint, the RSD-based algorithm gets trapped in local optima and fails to converge, whereas the RTR-SCA algorithm successfully finds the global optimum and converges. When the PEB constraint is relaxed to \( \Gamma = 0.4 \) m, the RTR-SCA algorithm continues to outperform the RSD-based algorithm, showing faster convergence and improved communication performance. This demonstrates the benefits introduced by the use of second-order information.

\begin{figure}[t] \vspace{-0.4cm} 
	\centerline{\includegraphics[width=3.4in]{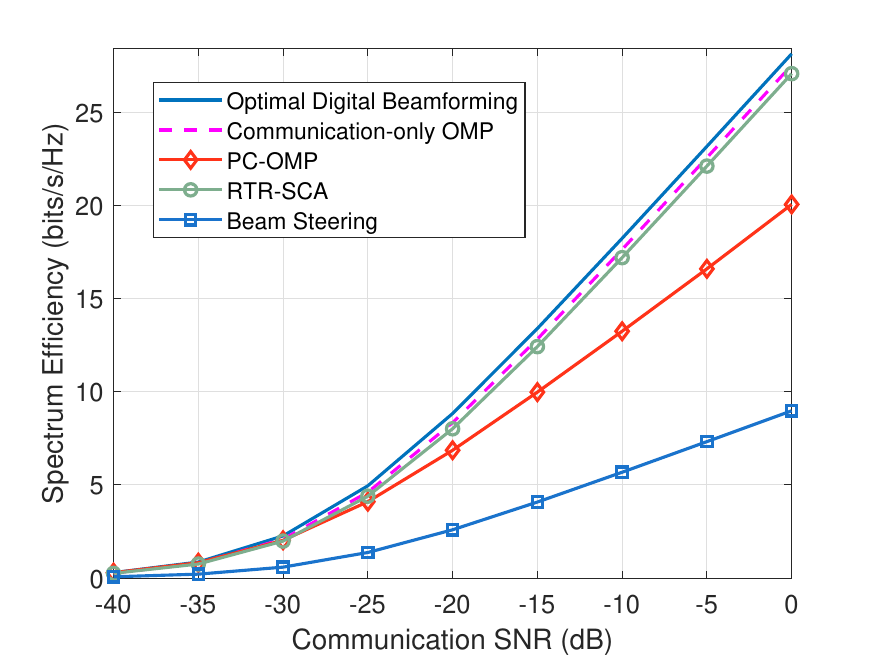}}
	\caption{SE versus $\text{SNR}_{\mathrm{c}}$ when $N=1$, $N_{\mathrm{s}}=N_{\mathrm{RF}}$\\$=3$ and $\Gamma=0.4$ m.}
	\label{fig:compare}
\end{figure}

We then analyze the SE achieved by the proposed algorithms with $N_{\mathrm{s}} = N_{\mathrm{RF}}$, which represents the worst-case scenario since we assume $N_{\mathrm{RF}} \geq N_{\mathrm{s}}$. Fig. \ref{fig:compare} illustrates the SE achieved by the RTR-SCA and PC-OMP design algorithms under different $\text{SNR}_\mathrm{c}$. For comparison, we introduce two benchmarks: the beam steering towards the channel's dominant physical direction and the communication-only OMP algorithm from \cite{spatially}. The results show that when the PEB requirement is 0.4 m, the RTR-SCA algorithm outperforms both the PC-OMP algorithm, although PC-OMP has significantly lower computational complexity. Also, the communication-only OMP achieves the highest SE, highlighting the degradation caused by the involvement of the positioning task. All algorithms exhibit better SE than beam steering.

\begin{figure}[t] \vspace{-0.4cm} 
	\centerline{\includegraphics[width=3.4in]{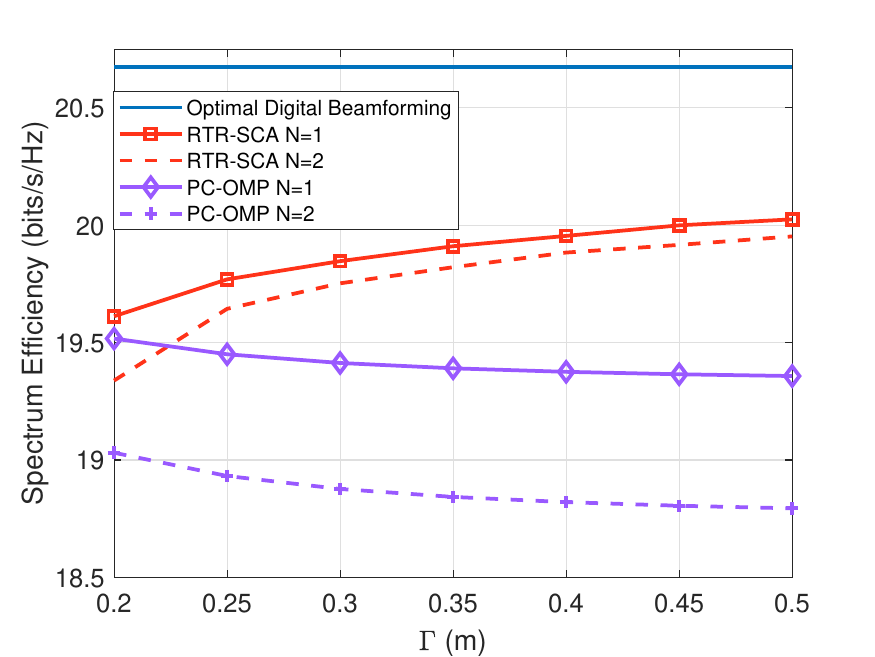}}
	\caption{SE versus PEB threshold $\Gamma$ when $N_{\mathrm{s}}=2$, $N_{\mathrm{RF}}=4$ and $\text{SNR}_\mathrm{c}=0$ dB.}
	\label{fig:tradeoff}
\end{figure}

Fig. \ref{fig:tradeoff} illustrates the impact of the PEB threshold $\Gamma$ on SE. The SE of the RTR-SCA algorithm improves as $\Gamma$ increases, highlighting the tradeoff between positioning and communication performance. The RTR-SCA algorithm is generally more sensitive to changes in $\Gamma$ because $\Gamma$ directly affects both the analog and digital beamformer designs. In contrast, the PC-OMP algorithm is less influenced by $\Gamma$, as it only affects a small portion of the total power allocated within the beamformer, resulting in minimal tradeoff for PC-OMP.
Under a strict PEB constraint, a slight deviation in SE is observed for the PC-OMP algorithm, suggesting that its primary limitation arises from the $N$ degree of freedom sacrificed in the beamformer design for communication, which increases $\delta$ in Lemma 2, thereby causing a minor deviation in the integrated beamformer's norm. Also, increasing $N$ from 1 to 2 requires power for different positioning points, which significantly degrades the communication performance.

\begin{figure}[t] \vspace{-0.4cm} 
	\centerline{\includegraphics[width=3.4in]{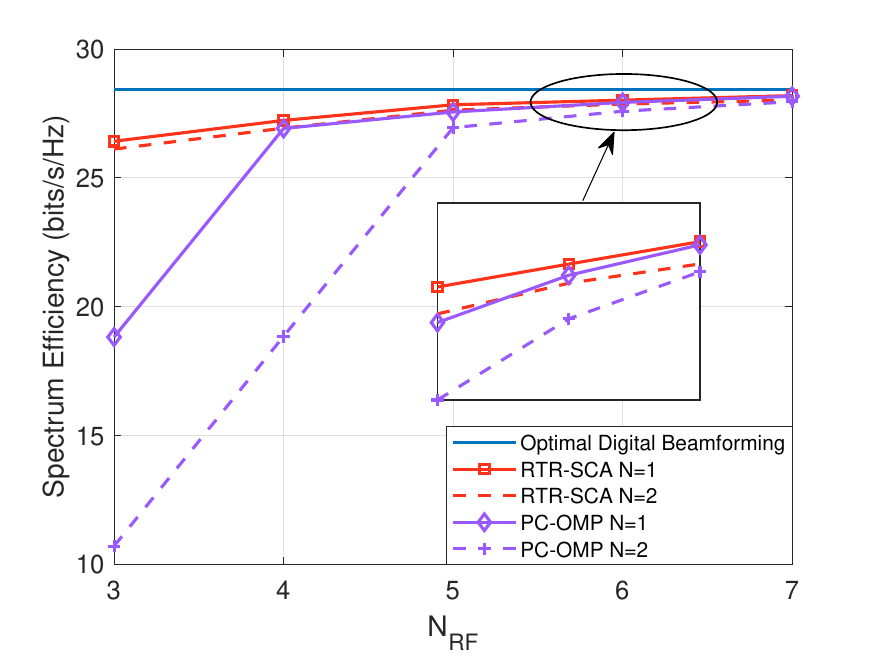}}
	\caption{SE versus $N_{\mathrm{RF}}$ when $N_{\mathrm{s}}=3$, $\Gamma=0.5$ m and $\text{SNR}_\mathrm{c}=0$ dB.}
	\label{fig:nrf}
\end{figure}

Finally, we fix $N_{\mathrm{s}} = 3$ and change $N_{\mathrm{RF}}$. Fig. \ref{fig:nrf} illustrates the impact of $N_{\mathrm{RF}}$ on communication performance with $N_{\mathrm{s}} = 3$ when a positioning accuracy of 0.5 m is required. In general, the RTR-SCA algorithm outperforms the PC-OMP algorithm, and the SE increases as $N_{\mathrm{RF}}$ grows. Notably, if $N_{\mathrm{RF}} < N_{\mathrm{s}} + N$, the performance of the PC-OMP algorithm is significantly degraded. This can be attributed to the limitations of the PC-OMP algorithm, as fixing $N$ columns of $\mathbf{F}_{\mathrm{RF}}$ reduces the degree of freedom for communication design. The harm is negligible when $N_{\mathrm{RF}} \geq N_{\mathrm{s}} + N$, at which point the performance improves substantially. Additionally, although \cite{yuwei} demonstrates that when $N_{\mathrm{RF}} \geq 2N_{\mathrm{s}}$, a closed-form optimal solution exists, yielding the same SE as the optimal digital beamformer, this theorem does not hold when the positioning task is incorporated, as shown by the points at $N_{\mathrm{RF}} \geq 2N_{\mathrm{s}}$ in Fig. \ref{fig:nrf}.

\section{Conclusion} \label{sec:conclusion}
In this study, we integrated positioning and communication in a mmWave ISAC MIMO-OFDM system.  A closed-form expression for the PEB was derived using OFDM signals for target positioning, and the hybrid beamformer was optimized based on this expression. Two efficient algorithms, named RTR-SCA and PC-OMP, were developed and thoroughly verified. simulation results provided valuable insights: First, while the proposed hybrid AOA/TOA positioning scheme mitigated the impact of channel gain uncertainty, targets along the line connecting the centers of the two BS arrays remained challenging to position. Second, the RTR-SCA method effectively addressed the issue of local optima and slow convergence in gradient-based methods. Also, the PC-OMP algorithm, as a low-complexity alternative to RTR-SCA, demonstrated strong performance with significantly reduced computational complexity.
For future work, considering more types of targets, including moving and extended ones, could extend the beamforming design. Also, scenarios involving multiple targets or BSs present potential avenues to expand our research.



\appendices

\section{Exact expressions of FIM elements} \label{appen:FIM}
Under mmWave conditions, as illustrated by \cite{henk}, the multipath components can be analyzed individually. Thus, we discuss the entries of the $l$-th submatrix $\mathbf{J}_{\boldsymbol{\xi}_l}$ below.

Before the derivation, we define the following notations,
\begin{subequations} \label{eq:short}
   \begin{align}
   \dot{\mathbf{a}}_{{\theta,l}}&\triangleq\frac{\partial\boldsymbol{\Lambda}_{{\mathrm{t}}}^{{\mathrm{T}}}\mathbf{k}(\theta^{\mathrm{t}}_l,\phi^{\mathrm{t}}_l)}{\partial\theta^{\mathrm{t}}_l}\odot\mathbf{a}_{l}\notag\\&=\boldsymbol{\Lambda}_{{\mathrm{t}}}^{{\mathrm{T}}}\dot{\mathbf{k}}_{\theta,l}\odot\mathbf{a}_{l},
   \\\dot{\mathbf{a}}_{{\phi,l}}&=\boldsymbol{\Lambda}_{{\mathrm{t}}}^{{\mathrm{T}}}\dot{\mathbf{k}}_{\phi,l}\odot\mathbf{a}_{l},
   \\\dot{\mathbf{b}}_{{\theta,l}}&=\boldsymbol{\Lambda}_{\mathrm{r}}^{\mathrm{T}}\dot{\mathbf{k}}_{\theta,l}\odot\mathbf{b}_{l}, \label{eq:shortb1}
   \\\dot{\mathbf{b}}_{{\phi,l}}&=\boldsymbol{\Lambda}_{\mathrm{r}}^{\mathrm{T}}\dot{\mathbf{k}}_{\phi,l}\odot\mathbf{b}_{l},\label{eq:shortb2}
   \end{align} 
\end{subequations}
with $\dot{\mathbf{k}}_{\theta,l}$ given by $\dot{\mathbf{k}}_{\theta,l}=\frac{2\pi}\lambda[\cos\theta\cos\phi,\cos\theta\sin\phi,-\sin\theta]^\mathrm{T}$ and $\dot{\mathbf{k}}_{\phi,l}$ by $\dot{\mathbf{k}}_{\phi,l}=\frac{2\pi}\lambda[-\sin\theta\sin\phi,\sin\theta\cos\phi,0]^\mathrm{T}$. Moreover, $\boldsymbol{\Lambda}_{{\mathrm{t}}}$ denotes the Cartesian coordinates of the elements in $\text{BS}_1$.

Based on \eqref{eq:FIM} and the definitions in \eqref{eq:short}, the entries of the $l$-th submatrix $\mathbf{J}_{\boldsymbol{\xi}_l}$ can be derived as
\begin{subequations}
   \begin{align}
   \mathbf{J}_{\boldsymbol{\xi}_l}\!(\theta^{\mathrm{r}}_l,\theta^{\mathrm{r}}_l)&=\gamma\Re\Big\{|\beta|^2\dot{\mathbf{b}}_{{\theta,l}}^{\mathrm{H}}\dot{\mathbf{b}}_{{\theta,l}}\sum\nolimits_{k=1}^K\!\mathbf{a}_l^{\mathrm{H}}\mathbf{F}_k\mathbf{F}_k^{\mathrm{H}}\mathbf{a}_l\Big\},\notag
   \\\mathbf{J}_{\boldsymbol{\xi}_l}\!(\theta^{\mathrm{r}}_l,\phi^{\mathrm{r}}_l)&=\gamma\Re\Big\{|\beta|^2\dot{\mathbf{b}}_{{\theta,l}}^{\mathrm{H}}\dot{\mathbf{b}}_{{\phi,l}}\sum\nolimits_{k=1}^K\!\mathbf{a}_l^{\mathrm{H}}\mathbf{F}_k\mathbf{F}_k^{\mathrm{H}}\mathbf{a}_l\Big\},\notag
   \\\mathbf{J}_{\boldsymbol{\xi}_l}\!(\theta^{\mathrm{r}}_l,\theta^{\mathrm{t}}_l)&=\gamma\Re\Big\{-|\beta|^2\dot{\mathbf{b}}_{{\theta,l}}^{\mathrm{H}}\mathbf{b}_l\sum\nolimits_{k=1}^K\!\dot{\mathbf{a}}_{{\theta,l}}^{\mathrm{H}}\mathbf{F}_k\mathbf{F}_k^{\mathrm{H}}\mathbf{a}_l\Big\}, \notag
   \\\mathbf{J}_{\boldsymbol{\xi}_l}\!(\theta^{\mathrm{r}}_l,\tau_l)&=\gamma\Re\Big\{|\beta|^2\dot{\mathbf{b}}_{{\theta,l}}^{\mathrm{H}}\mathbf{b}_l\sum\nolimits_{k=1}^K\!2\pi f_k\dot{\mathbf{a}}_{{\theta,l}}^{\mathrm{H}}\mathbf{F}_k\mathbf{F}_k^{\mathrm{H}}\mathbf{a}_l\Big\}, \notag
   \\\mathbf{J}_{\boldsymbol{\xi}_l}\!(\theta^{\mathrm{r}}_l,\beta_l^{\mathrm{R}})&=\gamma\Re\Big\{j\beta^{*}\dot{\mathbf{b}}_{{\theta,l}}^{\mathrm{H}}\mathbf{b}_l\sum\nolimits_{k=1}^K\!\dot{\mathbf{a}}_{{\theta,l}}^{\mathrm{H}}\mathbf{F}_k\mathbf{F}_k^{\mathrm{H}}\mathbf{a}_l\Big\}, \notag
   \\\mathbf{J}_{\boldsymbol{\xi}_l}\!(\theta^{\mathrm{r}}_l,\beta_l^{\mathrm{I}})&=\gamma\Re\Big\{-\beta^{*}\dot{\mathbf{b}}_{{\theta,l}}^{\mathrm{H}}\mathbf{b}_l\sum\nolimits_{k=1}^K\!\dot{\mathbf{a}}_{{\theta,l}}^{\mathrm{H}}\mathbf{F}_k\mathbf{F}_k^{\mathrm{H}}\mathbf{a}_l\Big\}, \notag
   \\\mathbf{J}_{\boldsymbol{\xi}_l}\!(\theta^{\mathrm{t}}_l,\theta^{\mathrm{t}}_l)&=\gamma\Re\Big\{|\beta|^2\mathbf{b}_l^{\mathrm{H}}\mathbf{b}_l\sum\nolimits_{k=1}^K\!\dot{\mathbf{a}}_{{\theta,l}}^{\mathrm{H}}\mathbf{F}_k\mathbf{F}_k^{\mathrm{H}}\dot{\mathbf{a}}_{{\theta,l}}\Big\}, \notag
   \\\mathbf{J}_{\boldsymbol{\xi}_l}\!(\theta^{\mathrm{t}}_l,\phi^{\mathrm{t}}_l)&=\gamma\Re\Big\{|\beta|^2\mathbf{b}_l^{\mathrm{H}}\mathbf{b}_l\sum\nolimits_{k=1}^K\!\dot{\mathbf{a}}_{{\phi,l}}^{\mathrm{H}}\mathbf{F}_k\mathbf{F}_k^{\mathrm{H}}\dot{\mathbf{a}}_{{\theta,l}}\Big\}, \notag
   \\\mathbf{J}_{\boldsymbol{\xi}_l}\!(\theta^{\mathrm{t}}_l,\tau_l)&=\gamma\Re\Big\{-|\beta|^2\mathbf{b}_l^{\mathrm{H}}\mathbf{b}_l\sum\nolimits_{k=1}^K\!2\pi f_k\mathbf{a}_l^{\mathrm{H}}\mathbf{F}_k\mathbf{F}_k^{\mathrm{H}}\dot{\mathbf{a}}_{{\theta,l}}\Big\}, \notag
   \\\mathbf{J}_{\boldsymbol{\xi}_l}\!(\theta^{\mathrm{t}}_l,\beta_l^{\mathrm{R}})&=\gamma\Re\Big\{-j\beta^{*}\mathbf{b}_l^{\mathrm{H}}\mathbf{b}_l\sum\nolimits_{k=1}^K\!\mathbf{a}_l^{\mathrm{H}}\mathbf{F}_k\mathbf{F}_k^{\mathrm{H}}\dot{\mathbf{a}}_{{\theta,l}}\Big\}, \notag
   \\\mathbf{J}_{\boldsymbol{\xi}_l}\!(\theta^{\mathrm{t}}_l,\beta_l^{\mathrm{I}})&=\gamma\Re\Big\{\beta^{*}\mathbf{b}_l^{\mathrm{H}}\mathbf{b}_l\sum\nolimits_{k=1}^K\!\mathbf{a}_l^{\mathrm{H}}\mathbf{F}_k\mathbf{F}_k^{\mathrm{H}}\dot{\mathbf{a}}_{{\theta,l}}\Big\}, \notag
   \\\mathbf{J}_{\boldsymbol{\xi}_l}\!(\tau_l,\beta_l^{\mathrm{R}})&=\gamma\Re\Big\{j\beta^{*}\mathbf{b}_l^{\mathrm{H}}\mathbf{b}_l\sum\nolimits_{k=1}^K\!2\pi f_k\mathbf{a}_l^{\mathrm{H}}\mathbf{F}_k\mathbf{F}_k^{\mathrm{H}}\mathbf{a}_l\Big\}, \notag
   \\\mathbf{J}_{\boldsymbol{\xi}_l}\!(\tau_l,\beta_l^{\mathrm{I}})&=\gamma\Re\Big\{-\beta^{*}\mathbf{b}_l^{\mathrm{H}}\mathbf{b}_l\sum\nolimits_{k=1}^K\!2\pi f_k\mathbf{a}_l^{\mathrm{H}}\mathbf{F}_k\mathbf{F}_k^{\mathrm{H}}\mathbf{a}_l\Big\}, \notag
   \\\mathbf{J}_{\boldsymbol{\xi}_l}\!(\tau_l,\tau_l)&=\gamma\Re\Big\{|\beta|^{2}\mathbf{b}_l^{\mathrm{H}}\mathbf{b}_l\sum\nolimits_{k=1}^K\!4\pi^2 f_k^2\mathbf{a}_l^{\mathrm{H}}\mathbf{F}_k\mathbf{F}_k^{\mathrm{H}}\mathbf{a}_l\Big\}, \notag
   \\\mathbf{J}_{\boldsymbol{\xi}_l}\!(\beta_l^{\mathrm{R}},\beta_l^{\mathrm{R}})&=\gamma\Re\Big\{\mathbf{b}_l^{\mathrm{H}}\mathbf{b}_l\sum\nolimits_{k=1}^K\!\mathbf{a}_l^{\mathrm{H}}\mathbf{F}_k\mathbf{F}_k^{\mathrm{H}}\mathbf{a}_l\Big\}, \notag
   \\\mathbf{J}_{\boldsymbol{\xi}_l}\!(\beta_l^{\mathrm{R}},\beta_l^{\mathrm{I}})&=\gamma\Re\Big\{j\mathbf{b}_l^{\mathrm{H}}\mathbf{b}_l\sum\nolimits_{k=1}^K\!\mathbf{a}_l^{\mathrm{H}}\mathbf{F}_k\mathbf{F}_k^{\mathrm{H}}\mathbf{a}_l\Big\}, \notag
   \\\mathbf{J}_{\boldsymbol{\xi}_l}\!(\beta_l^{\mathrm{I}},\beta_l^{\mathrm{I}})&=\gamma\Re\Big\{\mathbf{b}_l^{\mathrm{H}}\mathbf{b}_l\sum\nolimits_{k=1}^K\!\mathbf{a}_l^{\mathrm{H}}\mathbf{F}_k\mathbf{F}_k^{\mathrm{H}}\mathbf{a}_l\Big\}, \notag
   \end{align} 
\end{subequations}
where $\gamma=\frac{2N_{\mathrm{r}}^{\mathrm{s}}N_{\mathrm{t}}E_{\mathrm{0}}M}{\sigma_{\mathrm{s}}^2N_{\mathrm{s}}}$ is the SNR factor for sensing, and $\mathbf{F}_k=\mathbf{F}_{\mathrm{RF}}\mathbf{F}_{\mathrm{BB},k}$ denotes the beamforming matrix for the $k$-th subcarrier.

\section{Proof of the sub-matrix structure} \label{appen:approx}
To approximate the submatrix into the structure shown in Fig. \ref{fig:FIM}, we first demonstrate that the AOA and TOA are uncorrelated with other channel parameters. 

In two-dimensional cases, the derivative of the channel response vector $\mathbf{a}(\theta)$ w.r.t. $\theta$ is orthogonal to itself when the array is centered at the origin \cite{liufancrb}. This result can be directly extended to the three-dimensional case, where we have
\begin{equation} \vspace{-0.1cm}
    \dot{\mathbf{b}}_{{\star,l}}^{\mathrm{H}}\mathbf{b}_l=0,
\end{equation}
with $\star\in\{\theta,\phi\}$. Thus, the AOA, $\theta_l^{\mathrm{R}}$ and $\phi_l^{\mathrm{R}}$, turn out to be independent of other channel parameters.
We also have 
\begin{equation} \vspace{-0.1cm}
    \sum\nolimits_{k=1}^K\!2\pi f_k=\sum\nolimits_{k=1}^K\!2\pi(k-\frac{K+1}{2})\Delta f=0.
\end{equation}
To gain fundamental understandings, we assume that all subcarriers share the same beamformer \cite{yangjie} and prove that the TOA, $\tau_l$, is uncorrelated with other channel parameters, which is consistent with \cite{henk}, \cite{evaluation}.

Next, we focus on the diagonal structure of the $2 \times 2$ $\beta$ matrix $\mathbf{J}_{\boldsymbol{\beta_l}}$ and the $2 \times 2$ AOA matrix $\mathbf{J}_{\mathrm{AOA}}$, with
\begin{equation} \vspace{-0.1cm}
    \mathbf{J}_{\boldsymbol{\beta_l}}=\mathbf{J}_{\boldsymbol{\xi_l}}(6\!:\!7,6\!:\!7),\ \ \mathbf{J}_{\mathrm{AOA}}=\mathbf{J}_{\boldsymbol{\xi_l}}(1\!:\!2,1\!:\!2).
\end{equation}
The diagonal form of the $\mathbf{J}_{\boldsymbol{\beta_l}}$ is straightforward to prove, as $\mathbf{J}_{\boldsymbol{\xi}_l}(\beta_l^{\mathrm{R}}, \beta_l^{\mathrm{I}})$ is the real part of an imaginary number, which equals to zero. To approximate $\mathbf{J}_{\mathrm{AOA}}$ as diagonal, we introduce the following definition:

\textit{Definition 2:} Consider a square matrix $\mathbf{U}(x)$. If it can be decomposed into a diagonal matrix $\mathbf{U}_{\mathrm{d}}(x)$ and a hollow matrix $\mathbf{U}_{\mathrm{h}}(x)$, we say that $\mathbf{U}(x) = \mathbf{U}_{\mathrm{d}}(x) + \mathbf{U}_{\mathrm{h}}(x)$ is almost diagonal (AD) w.r.t. any parameter $x$ if
\begin{equation} \vspace{-0.1cm}
    \lim_{x\to\infty}\delta(\mathbf{U},x)\triangleq\lim_{\kappa_n\to\infty}\frac{\|\mathbf{U}_{\mathrm{s}}(x)\|_\mathrm{F}}{\|\mathbf{U}_{\mathrm{d}}(x)\|_\mathrm{F}}=0.
\end{equation}
To approximate $\mathbf{J}_{\mathrm{AOA}}$, Definition 2 is adopted and we have
\begin{equation} \vspace{-0.1cm}
    \delta^2(\mathbf{J}_{\mathrm{AOA}},N_{\mathrm{r}}^{\mathrm{s}})=\frac{\|\mathbf{J}_{\mathrm{s}}(N_{\mathrm{r}}^{\mathrm{s}})\|^2_\mathrm{F}}{\|\mathbf{J}_{\mathrm{d}}(N_{\mathrm{r}}^{\mathrm{s}})\|^2_\mathrm{F}}=\frac{2\dot{\mathbf{b}}_{{\theta,l}}^{\mathrm{H}}\dot{\mathbf{b}}_{{\phi,l}}}{\dot{\mathbf{b}}_{{\theta,l}}^{\mathrm{H}}\dot{\mathbf{b}}_{{\theta,l}}+\dot{\mathbf{b}}_{{\phi,l}}^{\mathrm{H}}\dot{\mathbf{b}}_{{\phi,l}}}.
\end{equation}
According to \eqref{eq:short}, the denominator can be written as
\begin{equation} \vspace{-0.1cm}
    \begin{aligned}
        \dot{\mathbf{b}}_{{\theta,l}}^{\mathrm{H}}\dot{\mathbf{b}}_{{\phi,l}}&\overset{(a)}{=}\dot{\mathbf{k}}_{\theta,l}^{\mathrm{T}}\boldsymbol{\Lambda}_{\mathrm{r}}\mathrm{diag}(\mathbf{b}_l^{\mathrm{H}})\mathrm{diag}(\mathbf{b}_l)\boldsymbol{\Lambda}_{\mathrm{r}}^{\mathrm{T}}\dot{\mathbf{k}}_{\phi,l}
        \\&=\frac{1}{N_{\mathrm{r}}^{\mathrm{s}}}\dot{\mathbf{k}}_{\theta,l}^{\mathrm{T}}\boldsymbol{\Lambda}_{\mathrm{r}}\boldsymbol{\Lambda}_{\mathrm{r}}^{\mathrm{T}}\dot{\mathbf{k}}_{\phi,l}
    \end{aligned}
\end{equation}
where (a) follows from applying $\mathbf{a}\odot\mathbf{b}=\mathrm{diag}(\mathbf{b})\times\mathbf{a}$. Since the receive array is centered at $(0,0,0)$, we derive
\begin{equation} \vspace{-0.1cm}
    \boldsymbol{\Lambda}_{\mathrm{r}}\boldsymbol{\Lambda}_{\mathrm{r}}^{\mathrm{T}}=\mathrm{diag}(\mathbf{u}),
\end{equation}
with
\begin{equation} \vspace{-0.1cm}
    \mathbf{u}=\left[\sum\nolimits_{n=1}^{N_{\mathrm{r}}^{\mathrm{s}}}\!x_{\mathrm{r},n}^2,\sum\nolimits_{n=1}^{N_{\mathrm{r}}^{\mathrm{s}}}\!y_{\mathrm{r},n}^2,\sum\nolimits_{n=1}^{N_{\mathrm{r}}^{\mathrm{s}}}\!z_{\mathrm{r},n}^2\right]^{\mathrm{T}},
\end{equation}
where $[x_{\mathrm{r},n},y_{\mathrm{r},n},z_{\mathrm{r},n}]^{\mathrm{T}}$ represents the Cartesian coordinates of the $n$-th receive antenna element. Thus, the denominator and the numerator can be respectively calculated by
\begin{subequations}
    \begin{align}
        \dot{\mathbf{b}}_{{\theta,l}}^{\mathrm{H}}\dot{\mathbf{b}}_{{\phi,l}}&=\frac{4\pi^2}{N_{\mathrm{r}}^{\mathrm{s}}\lambda^2}\sin\theta^{\mathrm{r}}_l\cos\theta^{\mathrm{r}}_l\sin\phi^{\mathrm{r}}_l\cos\phi^{\mathrm{r}}_l \notag
        \\&\quad\quad\quad\quad\quad\quad\quad\times(\sum\nolimits_{n=1}^{N_{\mathrm{r}}^{\mathrm{s}}}x_{\mathrm{r},n}^2-\sum\nolimits_{n=1}^{N_{\mathrm{r}}^{\mathrm{s}}}y_{\mathrm{r},n}^2),\notag
        \\\dot{\mathbf{b}}_{{\theta,l}}^{\mathrm{H}}\dot{\mathbf{b}}_{{\theta,l}}&=\frac{4\pi^2}{N_{\mathrm{r}}^{\mathrm{s}}\lambda^2}\Big(\cos^2\theta^{\mathrm{r}}_l\cos^2\phi^{\mathrm{r}}_l\sum\nolimits_{n=1}^{N_{\mathrm{r}}^{\mathrm{s}}}x_{\mathrm{r},n}^2 \notag
        \\&+\cos^2\theta^{\mathrm{r}}_l\sin^2\phi^{\mathrm{r}}_l\sum\nolimits_{n=1}^{N_{\mathrm{r}}^{\mathrm{s}}}y_{\mathrm{r},n}^2\notag
        +\sin^2\theta^{\mathrm{r}}_l\sum\nolimits_{n=1}^{N_{\mathrm{r}}^{\mathrm{s}}}z_{\mathrm{r},n}^2\Big),\notag
        \\\dot{\mathbf{b}}_{{\phi,l}}^{\mathrm{H}}\dot{\mathbf{b}}_{{\phi,l}}&=\frac{4\pi^2}{N_{\mathrm{r}}^{\mathrm{s}}\lambda^2}\Big(\sin^2\theta^{\mathrm{r}}_l\sin^2\phi^{\mathrm{r}}_l\sum\nolimits_{n=1}^{N_{\mathrm{r}}^{\mathrm{s}}}x_{\mathrm{r},n}^2 \notag
        \\&\quad\quad\quad\quad\quad\quad+\sin^2\theta^{\mathrm{r}}_l\cos^2\phi^{\mathrm{r}}_l\sum\nolimits_{n=1}^{N_{\mathrm{r}}^{\mathrm{s}}}y_{\mathrm{r},n}^2\Big).\notag
    \end{align}
\end{subequations}
For a large number of receive antennas, we deduce
\begin{equation} \vspace{-0.1cm}
    \lim_{N_{\mathrm{r}}^{\mathrm{s}}\to\infty}\delta^2(\mathbf{J}_{\mathrm{AOA}},N_{\mathrm{r}}^{\mathrm{s}})\approx0,
\end{equation}
which completes the proof.


\section{Proof of Proposition 1} \label{appen:proposition1}
Considering CRBs of $\boldsymbol{\eta}_l$, we first assume that each subcarrier contributes to the error bound equally by defining
\begin{equation} \vspace{-0.1cm} \label{eq:G}
    G = \mathbf{a}_l^{\mathrm{H}}\mathbf{F}_k\mathbf{F}_k^{\mathrm{H}}\mathbf{a}_l, \forall k.
\end{equation}
Thus, we can derive
\begin{subequations} \label{eq:CRBs}
    \begin{align}
        \mathrm{CRB}_{\boldsymbol{\theta}_l^{\mathrm{r}}}&\triangleq\frac{1}{\mathbf{J}^{\mathrm{e}}_{\boldsymbol{\eta}_l}(1,1)}=\frac{1}{\gamma |\beta|^2 \dot{\mathbf{b}}_{{\theta,l}}^{\mathrm{H}}\dot{\mathbf{b}}_{{\theta,l}}KG}
        \\\mathrm{CRB}_{\boldsymbol{\phi}_l^{\mathrm{r}}}&\triangleq\frac{1}{\mathbf{J}^{\mathrm{e}}_{\boldsymbol{\eta}_l}(2,2)}=\frac{1}{\gamma |\beta|^2 \dot{\mathbf{b}}_{{\phi,l}}^{\mathrm{H}}\dot{\mathbf{b}}_{{\phi,l}}KG}
        \\\mathrm{CRB}_{\tau_l}&\triangleq\frac{1}{\mathbf{J}^{\mathrm{e}}_{\boldsymbol{\eta}_l}(3,3)}=\frac{3}{\pi^2\gamma |\beta|^2 B^2KG} \label{eq:CRB3}
    \end{align}
\end{subequations}
Notably, \eqref{eq:CRB3} is derived based on ${\mathbf{b}}_{l}^{\mathrm{H}}{\mathbf{b}}_{l}=1$. 
To transform $\mathbf{J}^{\mathrm{e}}_{\boldsymbol{\eta}_l}$ into CRBs of positioning, we write 
\begin{equation} \vspace{-0.1cm} \label{eq:crb}
    \mathbf{C}_{\mathbf{p}_l}=\boldsymbol{\Upsilon}_l(\mathbf{J}^{\mathrm{e}}_{\boldsymbol{\eta}_l})^{-1}\boldsymbol{\Upsilon}_l^{\mathrm{T}},
\end{equation}
where $\boldsymbol{\Upsilon}_l$ is the Jacobian matrix defined by \cite{kay}
\begin{equation} \vspace{-0.1cm} \label{eq:jacobian}
    \boldsymbol{\Upsilon}_l\triangleq\frac{\partial\mathbf{p}_l}{\partial\boldsymbol{\eta}_l}=\begin{bmatrix}\frac{\partial x_l}{\partial\theta^{\mathrm{r}}_l}&\frac{\partial x_l}{\partial\phi^{\mathrm{r}}_l}&\frac{\partial x_l}{\tau_l}\\\frac{\partial y_l}{\partial\theta^{\mathrm{r}}_l}&\frac{\partial y_l}{\partial\phi^{\mathrm{r}}_l}&\frac{\partial y_l}{\tau_l}\\\frac{\partial z_l}{\partial\theta^{\mathrm{r}}_l}&\frac{\partial z_l}{\partial\phi^{\mathrm{r}}_l}&\frac{\partial z_l}{\tau_l}\end{bmatrix}, 
\end{equation}
with the entries written as
\begin{subequations} \label{eq:svector}
   \begin{align}&\frac{\partial x_l}{\partial\theta^{\mathrm{r}}_l}=(c^2\tau_l^2-D^2)c\tau_l\cos\theta^{\mathrm{r}}_l\cos\phi^{\mathrm{r}}_l, \notag
   \\&\frac{\partial y_l}{\partial\theta^{\mathrm{r}}_l}=(c^2\tau_l^2-D^2)c\tau_l\cos\theta^{\mathrm{r}}_l\sin\phi^{\mathrm{r}}_l,\notag
   \\&\frac{\partial z_l}{\partial\theta^{\mathrm{r}}_l}=(c^2\tau_l^2-D^2)(D\sin\phi^{\mathrm{r}}_l-c\tau_l\sin\theta^{\mathrm{r}}_l),\notag
   \\&\frac{\partial x_l}{\partial\phi^{\mathrm{r}}_l}=(c^2\tau_l^2-D^2)(D\sin^2\theta^{\mathrm{r}}_l-c\tau_l\sin\theta^{\mathrm{r}}_l\sin\phi^{\mathrm{r}}_l),\notag
   \\&\frac{\partial y_l}{\partial\phi^{\mathrm{r}}_l}=(c^2\tau_l^2-D^2)c\tau_l\sin\theta^{\mathrm{r}}_l\cos\phi^{\mathrm{r}}_l,\notag
   \\&\frac{\partial z_l}{\partial\phi^{\mathrm{r}}_l}=(c^2\tau_l^2-D^2)D\sin\theta^{\mathrm{r}}_l\cos\theta^{\mathrm{r}}_l\cos\phi^{\mathrm{r}}_l,\notag
   \\&\frac{\partial x_l}{\tau_l}=\upsilon\sin\theta^{\mathrm{r}}_l\cos\phi^{\mathrm{r}}_l,\ \frac{\partial y_l}{\tau_l}=\upsilon\sin\theta^{\mathrm{r}}_l\sin\phi^{\mathrm{r}}_l,\ \frac{\partial z_l}{\tau_l}=\upsilon\cos\theta^{\mathrm{r}}_l,\notag
   \end{align} 
\end{subequations}
where $\upsilon=c(c^2\tau_l^2-2c\tau_lD\sin\theta^{\mathrm{r}}_l\sin\phi^{\mathrm{r}}_l+D^2)$.
Given that $\mathbf{J}^{\mathrm{e}}_{\boldsymbol{\eta}_l}$ has been approximated into a diagonal matrix in \eqref{eq:diag}, we further express the SPEB based on \eqref{eq:crb} as
\begin{equation} \vspace{-0.1cm}
\begin{aligned}
    \!\!\!\mathrm{SPEB}\!&\triangleq\mathrm{tr}(\mathbf{C}_{\mathbf{p}_l})
    \\&=\frac{\|\Upsilon_l(:,1)\|^2}{\mathbf{J}^{\mathrm{e}}_{\boldsymbol{\eta}_l}(1,1)}+\frac{\|\Upsilon_l(:,2)\|^2}{\mathbf{J}^{\mathrm{e}}_{\boldsymbol{\eta}_l}(2,2)}+\frac{\|\Upsilon_l(:,3)\|^2}{\mathbf{J}^{\mathrm{e}}_{\boldsymbol{\eta}_l}(3,3)}
    \\&=\frac{1}{4\omega^4}(o\times\mathrm{CRB}_{\boldsymbol{\theta}_l^{\mathrm{r}}} \!+\!p\times\mathrm{CRB}_{\boldsymbol{\phi}_l^{\mathrm{r}}} \!+\!q\times\mathrm{CRB}_{\tau_l}),
\end{aligned}
\end{equation}
where
\begin{subequations} \label{eq:coefficient}
   \begin{align}
   \omega=&c\tau_l-D\sin\theta^{\mathrm{r}}_l\sin\phi^{\mathrm{r}}_l,
   \\o=&(c^2\tau_l^2-D^2)^2(c^2\tau_l^2+D^2\sin^2\phi^{\mathrm{r}}_l-2Dc\tau_l\sin\theta^{\mathrm{r}}_l\sin\phi^{\mathrm{r}}_l),\label{eq:o}
   \\p=&(c^2\tau_l^2-D^2)^2(D^2\sin^4\theta^{\mathrm{r}}_l-2Dc\tau_l\sin^3\theta^{\mathrm{r}}_l\sin\phi^{\mathrm{r}}_l \notag
\\&\ \ \ \ +c^2\tau_l^2\sin^2\theta^{\mathrm{r}}_l+D^2\sin^2\theta^{\mathrm{r}}_l\cos^2\theta^{\mathrm{r}}_l\cos^2\phi^{\mathrm{r}}_l),\label{eq:p}
   \\q=&\upsilon^2.
   \end{align} 
\end{subequations}
\vspace{-0.8cm} 
\section{Proof of Lemma 2} \label{appen:norm}
For the analog beamformer $\mathbf{F}_{\mathrm{RF}}=[\mathbf{A}_{\mathrm{s}},\mathbf{F}_{\mathrm{RF}}^{\mathrm{ex}}]$ and the digital beamformer $\mathbf{F}_{\mathrm{BB},k}=[\mathbf{F}_{\mathrm{s},k},\mathbf{F}_{\mathrm{BB},k}^{\mathrm{ex}\,\mathrm{T}}]^\mathrm{T}$, we have 
\begin{equation} \vspace{-0.1cm}
\begin{aligned}
    \|\mathbf{F}_{\mathrm{RF}}\mathbf{F}_{\mathrm{BB},k}\|_F&=\|\mathbf{A}_{\mathrm{s}}\mathbf{F}_{\mathrm{s},k}^\mathrm{T}+\mathbf{F}_{\mathrm{RF}}^{\mathrm{ex}}\mathbf{F}_{\mathrm{BB},k}^{\mathrm{ex}}\|_F
    \\&=\|\mathbf{F}_{\mathrm{opt},k}^{\mathrm{ex}}\!+\!\mathbf{A}_{\mathrm{s}}\mathbf{F}_{\mathrm{s},k}^\mathrm{T}\!-\!(\mathbf{F}_{\mathrm{opt},k}^{\mathrm{ex}}\!-\!\mathbf{F}_{\mathrm{RF}}^{\mathrm{ex}}\mathbf{F}_{\mathrm{BB},k}^{\mathrm{ex}})\|_F
    \\&=\|\mathbf{F}_{\mathrm{opt},k}-(\mathbf{F}_{\mathrm{opt},k}^{\mathrm{ex}}-\mathbf{F}_{\mathrm{RF}}^{\mathrm{ex}}\mathbf{F}_{\mathrm{BB},k}^{\mathrm{ex}})\|_F. \notag
\end{aligned}
\end{equation}
As illustrated in \cite{alternating}, if we ignore the power constraint and approximate $\mathbf{F}_{\mathrm{opt},k}^{\mathrm{ex}}$ with the Euclidean distance satisfying $\left\|\mathbf{F}_{\mathrm{opt},k}^{\mathrm{ex}}-\mathbf{F}_\mathrm{RF}^{\mathrm{ex}}\hat{\mathbf{F}}_{\mathrm{BB},k}^{\mathrm{ex}}\right\|_F\leq\delta$, we have $\left\|\mathbf{F}_{\mathrm{opt},k}^{\mathrm{ex}}-\mathbf{F}_\mathrm{RF}^{\mathrm{ex}}\mathbf{F}_{\mathrm{BB},k}^{\mathrm{ex}}\right\|_F\leq2\delta$ after the normalization of $\hat{\mathbf{F}}_{\mathrm{BB},k}^{\mathrm{ex}}$. Thus, based on norm inequality, we further have
\begin{equation} \vspace{-0.1cm}
    \begin{aligned}
        \|\mathbf{F}_{\mathrm{opt},k}-(&\mathbf{F}_{\mathrm{opt},k}^{\mathrm{ex}}-\mathbf{F}_{\mathrm{RF}}^{\mathrm{ex}}\mathbf{F}_{\mathrm{BB},k}^{\mathrm{ex}})\|_F
        \\&\leq\|\mathbf{F}_{\mathrm{opt},k}\|_F+\|\mathbf{F}_{\mathrm{opt},k}^{\mathrm{ex}}-\mathbf{F}_{\mathrm{RF}}^{\mathrm{ex}}\mathbf{F}_{\mathrm{BB},k}^{\mathrm{ex}}\|_F
        \\&\leq\sqrt{N_\mathrm{s}}+2\delta,
    \end{aligned}
\end{equation}
and
\begin{equation} \vspace{-0.1cm}
    \begin{aligned}
        \|\mathbf{F}_{\mathrm{opt},k}-(&\mathbf{F}_{\mathrm{opt},k}^{\mathrm{ex}}-\mathbf{F}_{\mathrm{RF}}^{\mathrm{ex}}\mathbf{F}_{\mathrm{BB},k}^{\mathrm{ex}})\|_F
        \\&\geq|\|\mathbf{F}_{\mathrm{opt},k}\|_F-\|\mathbf{F}_{\mathrm{opt},k}^{\mathrm{ex}}-\mathbf{F}_{\mathrm{RF}}^{\mathrm{ex}}\mathbf{F}_{\mathrm{BB},k}^{\mathrm{ex}}\|_F|
        \\&\geq\sqrt{N_\mathrm{s}}-2\delta,
    \end{aligned}
\end{equation}
which indicates that $|\|\mathbf{F}_{\mathrm{RF}}\mathbf{F}_{\mathrm{BB},k}\|_F-\sqrt{N_{\mathrm{s}}}|\leq2\delta$, completing the proof.

\newpage

 




\vfill

\end{document}